\documentclass[a4paper,12pt]{article}
\usepackage[colorlinks=true,linkcolor=blue,urlcolor=blue,filecolor=black,citecolor=red,pdfstartview=FitV,pdftitle={},pdfsubject={},
pdfkeywords={},pdfpagemode=None,bookmarksopen=true]{hyperref}
\usepackage{graphicx}
\usepackage{epstopdf}%
\usepackage{amsmath}
\usepackage{amsfonts}
\usepackage{amssymb}
\usepackage{color}%
\usepackage{dcolumn}
\usepackage{slashed}
\usepackage{amssymb,ulem}
\usepackage{float}
\usepackage{amsthm,amsmath,amssymb}
\usepackage{mathrsfs}
\usepackage{subfigure}
\usepackage{wrapfig}
\usepackage{indentfirst}
\usepackage{multirow}
\usepackage{appendix}
\usepackage{graphicx} 
\usepackage{subfigure} 
\usepackage{tablefootnote}
\usepackage{suffix}
\usepackage{comment}
\WithSuffix\newcommand\footnote*[1]{\footnote[#1]{\addtocounter{footnote}{-1}}}

\usepackage{color}
\usepackage[dvipsnames]{xcolor}
\usepackage{cite}

\makeatletter
\renewcommand\@makefnmark{\hbox{\@textsuperscript{\normalfont\@thefnmark}}}

\newcommand{\Rmnum}[1]{\expandafter\@slowromancap\romannumeral #1@}

\makeatother

\begin{document}

\vspace{9mm}

\begin{center}
{{{\Large {\bf Existence of nonlinearly scalarized black holes in Einstein-scalar-Gauss-Bonnet theory with polynomial couplings }}}}\\[8mm]

{De-Cheng Zou$^{1}$\footnote{dczou@jxnu.edu.cn;}, Xu Yang$^{1}$, Meng-Yun Lai$^{1}$\footnote{ mengyunlai@jxnu.edu.cn}, Hyat Huang$^{1}$\footnote{hyat@mail.bnu.edu.cn}, Bo Liu$^{2}$\footnote{fenxiao2001@163.com},\\  Jutta Kunz$^{3}$\footnote{jutta.kunz@uni-oldenburg.de}, Yun Soo Myung$^{4}$\footnote{ysmyung@inje.ac.kr} and Rui-Hong Yue$^{5}$\footnote{rhyue@yzu.edu.cn}  }\\[8mm]

{
${}^1$College of Physics and Communication Electronics, Jiangxi Normal University, Nanchang 330022, China\\}
{${}^2$School of Arts and Sciences, Shaanxi University of Science and Technology, Xi'an, 710021, China\\}
{${}^3$Institut f\"ur  Physik, Universit\"at Oldenburg, Postfach 2503, D-26111 Oldenburg, Germany\\}
{${}^4$Center for Quantum Spacetime, Sogang University, Seoul 04107, Republic of  Korea\\}
{${}^5$Center for Gravitation and Cosmology, College of Physical Science and Technology, Yangzhou University, Yangzhou 225009, China}

\end{center}

\begin{abstract}

Nonlinearly scalarized black holes are 
investigated in Einstein-scalar-Gauss-Bonnet (EsGB) theory  with polynomial coupling functions $\zeta(\phi)$ satisfying $\zeta''(0) = 0$, where $\zeta'(\phi) = 0$ features besides $\phi=0$ solutions with constant $\phi_{\rm s} \ne 0$.
We determine the threshold amplitudes for Gaussian pulses, above which Schwarzschild black holes (SBHs) 
transition to scalarized black holes for two coupling functions: $\zeta(\phi)=\alpha\phi^4-\beta\phi^8$ and $\zeta(\phi)=\alpha\phi^4-\beta\phi^6$. 
In contrast, for the quartic coupling function $\zeta(\phi)=\alpha\phi^4$ SBHs are stable.
Treating $\zeta(\phi)R_{GB}^2$ as an effective potential $V_\text{eff}$ provides an explanation for the ``plateau" and the divergence observed in the time evolution. 
We then construct the branches of nonlinearly scalarized black holes in the probe limit and with backreaction.
While the pattern of the solution branches in the probe limit exhibits universal features, the presence of backreaction reveals a distinct dependence on the coupling strength $\beta$.

\end{abstract}

\section{Introduction}

Scalar fields appear ubiquitously in various contexts of theoretical physics.
For instance, they offer a straightforward and versatile framework that can account for the accelerating  universe, both in its early and later stages. 
On the other hand, the no-hair theorem seems to rule out black holes with scalar hair in Einstein theory, minimally coupled to a scalar field \cite{Bekenstein:1974sf}-\cite{Bekenstein:1995un}.  
However, this theorem can be circumvented in various ways \cite{Herdeiro:2015waa}.
Moreover, scalar hair may be present in theories that possess a non-minimal coupling of the scalar field to curvature or matter fields.  
In this direction, Damour and Esposito-Farese~\cite{Damour:1993hw} found the mechanism of spontaneous scalarization in scalar-tensor theories long ago, when studying neutron stars.

Among the scalar-tensor theories, Einstein-scalar-Gauss-Bonnet (EsGB) theory has received much attention in recent years.  
The advantages of this theory are multiple. 
First, it corresponds to a simple modification of Einstein gravity obtained by introducing a scalar coupling to the Gauss-Bonnet (GB) term (${\mathcal{R}^2}_{\rm GB}$). 
Second, it leads to second-order field equations since the GB term is topological in four dimensions.
Moreover, as shown in \cite{Doneva:2017bvd}-
\cite{Antoniou:2017hxj}, curvature induced spontaneous scalarization of Schwarzschild black holes (SBHs) may take place in EsGB theory when a tachyonic instability triggers the onset of scalarization.  
This mechanism is thus analogous to the matter induced spontaneous scalarization of neutron stars that is triggered by the coupling of a scalar field to matter.
Besides the fundamental branch ($n=0$) of scalarized black holes, excited branches of scalarized black holes arise, where the scalar field has $n$ radial nodes.
These branches of scalarized black holes bifurcate successively from the SBHs at critical values.
Spontaneous scalarization arises not only for the static SBHs but also for the rotating Kerr black holes (see e.g.~\cite{Antoniou:2017acq}-\cite{Berti:2020kgk}).
Interestingly, spontaneous scalarization of black holes is also observed when the scalar field is coupled to the Maxwell invariant.
In the resulting Einstein-Maxwell-scalar theory, spontaneous scalarization of Reissner-Nordstr\"om and Kerr-Newman black holes was observed \cite{Herdeiro:2018wub}-\cite{Promsiri:2023yda}.
Again, besides the fundamental spontaneously scalarized black holes, excited scalarized black holes were found. 

On the other hand, there exist further mechanisms for scalarization.
In the case of nonlinear scalarization of SBHs in EsGB theory, the effective mass term 
in the scalar equation always vanishes.
Nonlinear scalarization has, for instance, been achieved by employing exponential coupling functions of the form
\cite{Doneva:2021tvn}
\begin{eqnarray}
\zeta_{\rm ex}(\phi)=\frac{1}{4\kappa}\left(1-e^{-\kappa\phi^4}\right) \quad {\rm or} \quad
\frac{1}{4\kappa}\left(1-e^{-\kappa\phi^6}\right) ,
\label{cpin}
\end{eqnarray}
with coupling constant $\kappa$.
SBHs are then stable under linear scalar perturbations because the tachyonic instability is absent. 
Nevertheless, it is worth noting that SBHs become unstable against nonlinear scalar perturbations when the amplitude $A$ of the scalar perturbation exceeds a threshold amplitude $A_{\rm th}$ \cite{Doneva:2021tvn}-
\cite{Belkhadria:2023ooc}. 
In this case, 
branches of nonlinearly scalarized black holes were found, whose properties depend on the chosen coupling function and coupling constant.
A radial mode analysis of the nonlinearly scalarized solutions was performed in \cite{Blazquez-Salcedo:2022omw}, revealing stable and unstable branches and the loss of hyperbolicity on some of these branches when varying $\kappa$.

Reference \cite{Pombo:2023lxg} discussed the influence of a scalar field mass and self-interaction on the existence of the scalarized phases and the presence of jumps between 
bald and hairy black holes. 
Similarly, nonlinear scalarization was found for multi-scalar Gauss-Bonnet black holes \cite{Staykov:2024jbq,Staykov:2022uwq} and Schwarzschild black holes in scalar-torsion teleparallel gravity \cite{Gonzalez:2024ifp}. 
Furthermore, when evolving the nonlinear scalar equation in time on the Kerr black hole background, a threshold amplitude of the scalar perturbation was found, above which the Kerr black hole 
scalarizes \cite{Doneva:2022yqu}, 
employing exponential coupling functions in EsGB theory.  
Later, Lai \textit{et al.} \cite{Lai:2023gwe} explicitly constructed three branches of rotating nonlinearly scalarized black holes, choosing the quartic coupling function in Eq.~\eqref{cpin} 
using the pseudo-spectral method. 
Recently, rotating nonlinearly scalarized black holes were constructed within an EsGB framework that includes an additional squared GB term \cite{Liu:2025eve}. 

In the present work, we wish to explore nonlinear scalarization in EsGB theory by adopting polynomial coupling functions $\zeta(\phi)$  
instead of exponential ones. 
These coupling functions allow for Schwarzschild solutions with a constant scalar field that can vanish or have a finite value which is determined by the polynomial coefficients.
In particular, we introduce an effective potential for the scalar field and relate its properties to the presence of scalarized solutions.

The paper is organized as follows: 
In section \ref{Sec2}, we introduce EsGB theory and numerically solve the nonlinear scalar equation to investigate the scalarization 
of SBHs under Gaussian pulse perturbations. 
We then introduce an effective potential and investigate its properties for these polynomial couplings. 
In section \ref{Sec3}, we construct the branches of scalarized black hole solutions for the polynomial couplings.
First we consider the probe limit and then we solve the full set of coupled equations. %
We also compare our results to those of with those of exponential coupling functions.
We present our conclusions in section \ref{Sec5}.

\section{Nonlinear instability of SBHs in EsGB theory}
\label{Sec2}

\subsection{EsGB theory}

We start with EsGB theory, whose action is given by \cite{Doneva:2017bvd}
\begin{equation}
\mathcal{S} = \frac{1}{{16\pi }}\int {{d^4}} x\sqrt { - g} \left[ {R-2{\nabla _\mu }\phi {\nabla ^\mu }\phi +\lambda^2\zeta(\phi ){\mathcal{R}^2}_{\rm GB}} \right] , \label{action}
\end{equation}
where $R$ is the Ricci scalar, $\lambda$ is the coupling constant with dimension of length, and ${{\mathcal R}^2}_{\rm GB} = {R_{\mu \nu \rho \sigma }}{R^{\mu \nu \rho \sigma }}-4{R_{\mu\nu }}{R^{\mu \nu }} + {R^2}$ represents the GB term. Moreover, $\phi$ denotes the scalar field and  $\zeta(\phi)$ represents the scalar coupling function.  

Then, the field equations can be obtained as
\begin{eqnarray}
&&{R_{\mu \nu }} - \frac{1}{2}R{g_{\mu \nu }} + {\Gamma _{\mu \nu }} = 2{\nabla _\mu }\phi {\nabla _\nu }\phi  -{g_{\mu \nu }}{\nabla _\alpha }\phi {\nabla ^a}\phi,\label{field1}\\
&&\square \phi+\frac{\lambda^2}{4}\frac{{d\zeta(\phi )}}{{d\phi }}{\mathcal{R}^2}_{\rm GB}=0,\label{KG1}
\end{eqnarray}
where  $\Gamma _{\mu \nu}$ is defined by
\begin{eqnarray}
{\Gamma_{\mu\nu }}& =& - R({\nabla_\mu }{\psi_\nu} + {\nabla_\nu }{\psi _\mu })-4{\nabla ^\alpha }{\psi_\alpha }\left( {{R_{\mu\nu }}-\frac{1}{2}R{g_{\mu \nu }}}\right)+4{R_{\mu \alpha }}{\nabla ^\alpha }{\psi_\nu}\nonumber\\
&&+ 4{R_{\nu \alpha}}{\nabla^\alpha }{\psi_\mu}-4{g_{\mu\nu}}{R^{\alpha \beta }}{\nabla_\alpha }{\psi_\beta}+4{R^\beta}_{\mu \alpha\nu}\nabla_\alpha {\psi_\beta }\label{field3}
\end{eqnarray}
with
\begin{eqnarray}
{\psi _\mu } =\lambda^2\frac{{d\zeta(\phi )}}{{d\phi }}{\nabla _\mu }\phi. \nonumber
\end{eqnarray}

We recall that when the condition 
\begin{equation}
   \left.  \frac{{d\zeta(\phi )}}{{d\phi }} \right|_{\phi_{\rm s}}=0
   \label{const}
\end{equation}
holds, the field equations allow for the Schwarzschild solution with a constant scalar field $\phi_{\rm s}$.
Typically, the coupling functions are chosen so that $\phi=0$ is a solution.
However, in the case of polynomial coupling functions, additional solutions $\phi_{\rm s}$ of \eqref{const} may exist \cite{Silva:2018qhn}.

\subsection{Scalarization}

In EsGB theory, the linearized Einstein equation [$\delta G_{\mu\nu}(h)=0$] governing the metric perturbation $\delta g_{\mu\nu}=h_{\mu\nu}$ in the Schwarzschild background is  decoupled from the linearized scalar equation.  
Then, the linearized Einstein equation is the same as that for Einstein gravity.  
Since the SBH is stable against metric perturbations $h_{\mu\nu}$, we now focus on considering the linearized scalar equation
\begin{eqnarray}
 \bar{\square} \delta\phi-\mu_{{\rm{eff}}}^2\delta \phi = 0 ,
\end{eqnarray}
where the effective scalar mass  is given by
\begin{eqnarray}
\mu_{{\rm{eff}}}^2= -\frac{\lambda^2}{4}\frac{{{d^2}\zeta}}{{d{\phi ^2}}}(0)\bar{R}_{\rm GB}^2.
\end{eqnarray}
A tachyonic instability of the SBH may be triggered by an effective scalar mass $\mu^2_{{\rm{eff}}}$ when the coupling function $\zeta(\phi)$ satisfies the conditions
\begin{eqnarray}
\zeta(0) = 0, \quad \frac{{d\zeta}}{{d\phi }}(0) = 0, \quad \frac{{{d^2}\zeta}}{{d{\phi ^2}}}(0)  \ne  0.
\end{eqnarray}
Then spontaneous scalarization may take place around SBHs when the curvature is sufficiently large.  
In that case, a countable number of branches of scalarized black holes (possessing $n$ nodes of the scalar field, $n=0,~1, ~2, ~\dots $) emerge from scalar clouds, obtained as solutions of the static linearized scalar equation [$(\bar{\nabla}^2 -\mu_{{\rm{eff}}}^2)\delta \phi(\bf{x})=0$] 
\cite{Doneva:2017bvd}-\cite{Antoniou:2017acq}.

If the coupling function takes the exponential form given in Eq~(\ref{cpin}), the effective mass vanishes, however,
\begin{eqnarray}
\quad\frac{{{d^2}\zeta}}{{d{\phi ^2}}}(0)  =  0 \to \mu_{{\rm{eff}}}^2=0.\label{cond1}
\end{eqnarray}
Then it is clear that the tachyonic instability is absent, $\bar{\square}\delta \phi=0$, and the SBH is stable against linear scalar perturbations.
Nevertheless, the SBH may become unstable against nonlinear scalar perturbations when the scalar amplitude exceeds a threshold amplitude \cite{Doneva:2021tvn,Zhang:2023jei,Pombo:2023lxg}.
This provides a mechanism to obtain 
scalarized black holes via nonlinear scalarization.

\subsection{Time evolution}

To explore the nonlinear instability explicitly, we assume coupling functions satisfying 
\begin{eqnarray}
\frac{{{d^2}\zeta}}{{d{\phi ^2}}}(0)  = 0,
\end{eqnarray}
and study the time evolution of the Klein-Gordon equation on the background SBH spacetime
\begin{eqnarray} \label{SBH}
   ds^2_{\rm SBH}&=&-f(r)dt^2+f(r)^{-1}dr^2+r^2d\Omega^2_2,\quad f(r)=1-\frac{2M}{r}. 
\end{eqnarray}
The ($1+1$)-dimensional evolution of the scalar equation \eqref{KG1} can be written for $\Psi(t,r)=r \phi$ as
\begin{eqnarray}
-\frac{\partial^2\Psi(t,x)}{\partial{t}^2}&+&\frac{\partial^2\Psi(t,x)}{\partial{x}^2}-\frac{f(r) f'(r) }{r}\Psi(t,x)\nonumber\\
&&+\frac{\lambda^2f(r)\left[f(r) f''(r) - f''(r)+f'(r)^2\right]}{r}\frac{{d\zeta(\phi )}}{{d\phi }}\Psi(t,x)=0 
\label{KG2}
\end{eqnarray}
with the tortoise coordinate $x=\int f^{-1}(r)dr$. 

To solve the above equation numerically in the time domain, we employ finite differences of 7th order to approximate the spatial derivatives and perform the time integration using the 4th order Runge-Kutta method.
The boundary conditions we have to impose when evolving in time are that the scalar field takes the form of an outgoing wave at infinity and an ingoing wave at the black hole horizon.  
Note that we usually choose the computational domain in the radial direction to span the interval $[r_H+\delta r, r_\infty]$. 
Here $\delta r$ is a small shift away from the outer horizon radius $r_H$, where the coordinate system would become singular, and $r_\infty$ denotes the radius of the grid at its outer edge. 
When introducing the tortoise coordinate $x$, this interval is mapped to the tortoise coordinate domain $[x_{-\infty}, x_{+\infty}]$. 

Here we choose the initial state of the scalar field to be a Gaussian distribution with $A$ an initial amplitude 
\begin{eqnarray}
\Psi(t=0,x)=A e^{-\frac{(x-x_c)^2}{2\sigma^2}},
\end{eqnarray}
localized outside the horizon with the maximum at $x_c=-84.82$ and with the width $\sigma=1$. 
For the tortoise coordinate $x$, we choose a grid that spans the interval $ [-120,200]$. 
The von Neumann stability condition can be fulfilled when confining the calculations to an appropriate spatial region while keeping a small ratio of $\Delta t \leq \Delta x $. 
Accordingly, we take $\Delta t=0.1$ and $\Delta x=0.125$ in our calculations.

Making use of Eq.~\eqref{KG2}, we discuss the time evolution of the scalar field on the SBH background by choosing three polynomial couplings 
\begin{eqnarray}
\zeta_1(\phi) = \alpha {\phi ^4} - \beta {\phi ^8},\quad \zeta_2(\phi)=\alpha {\phi ^4} - \beta {\phi ^6},\quad \zeta_3(\phi)=\alpha\phi^4. 
\label{coupf}
\end{eqnarray}
Following \cite{Doneva:2021tvn}, we place a virtual
observer at a fixed location $r_\text{ob}$ to monitor the evolution of $\phi(r_\text{ob},t)$. 
When the signal $\phi(r_\text{ob},t)$ stabilizes, this indicates the onset of nonlinear scalarization.

\begin{figure}[H]
\centering
\subfigure[$\zeta_1(\phi)=\alpha\phi^4-\beta\phi^8$]{
\label{time1} 
\includegraphics[width=0.3\textwidth]{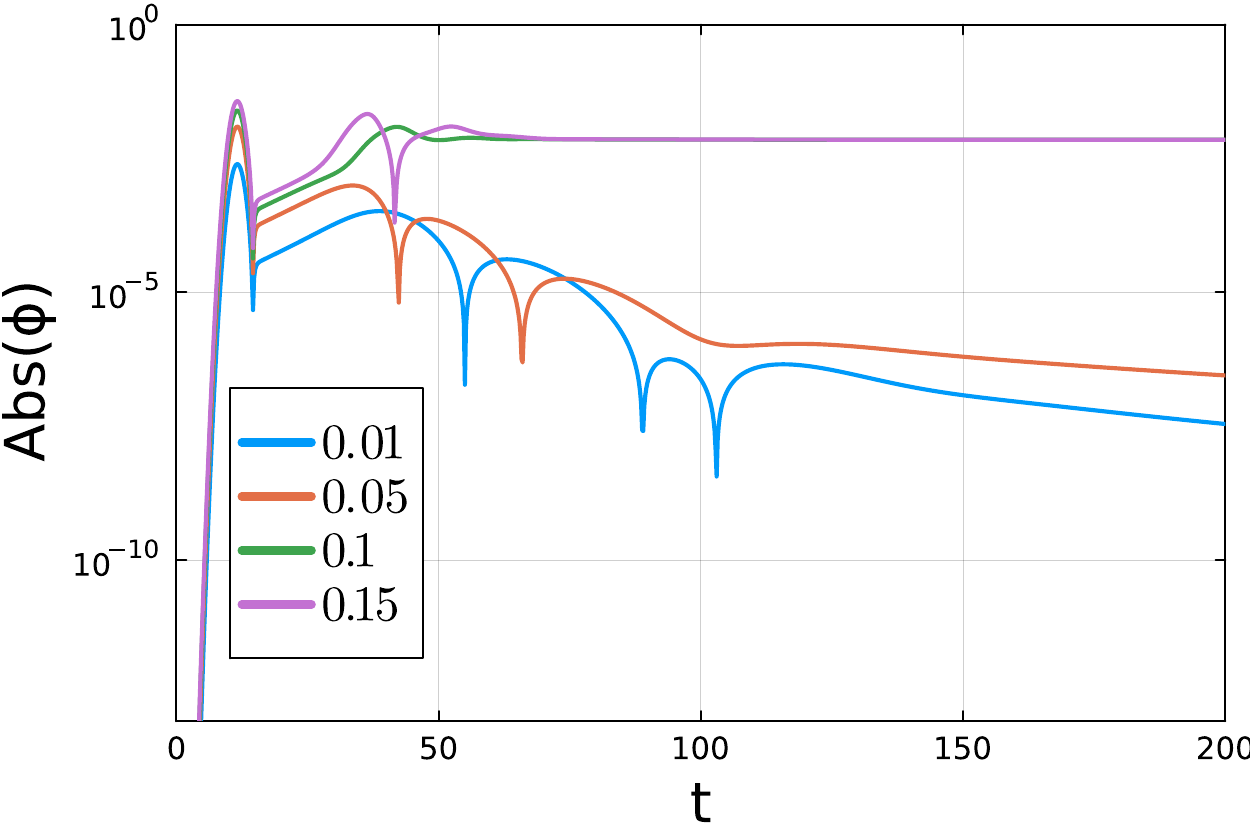}}
\hfill%
\subfigure[$\zeta_2(\phi)=\alpha\phi^4-\beta\phi^6$]{
\label{time2} 
\includegraphics[width=0.3\textwidth]{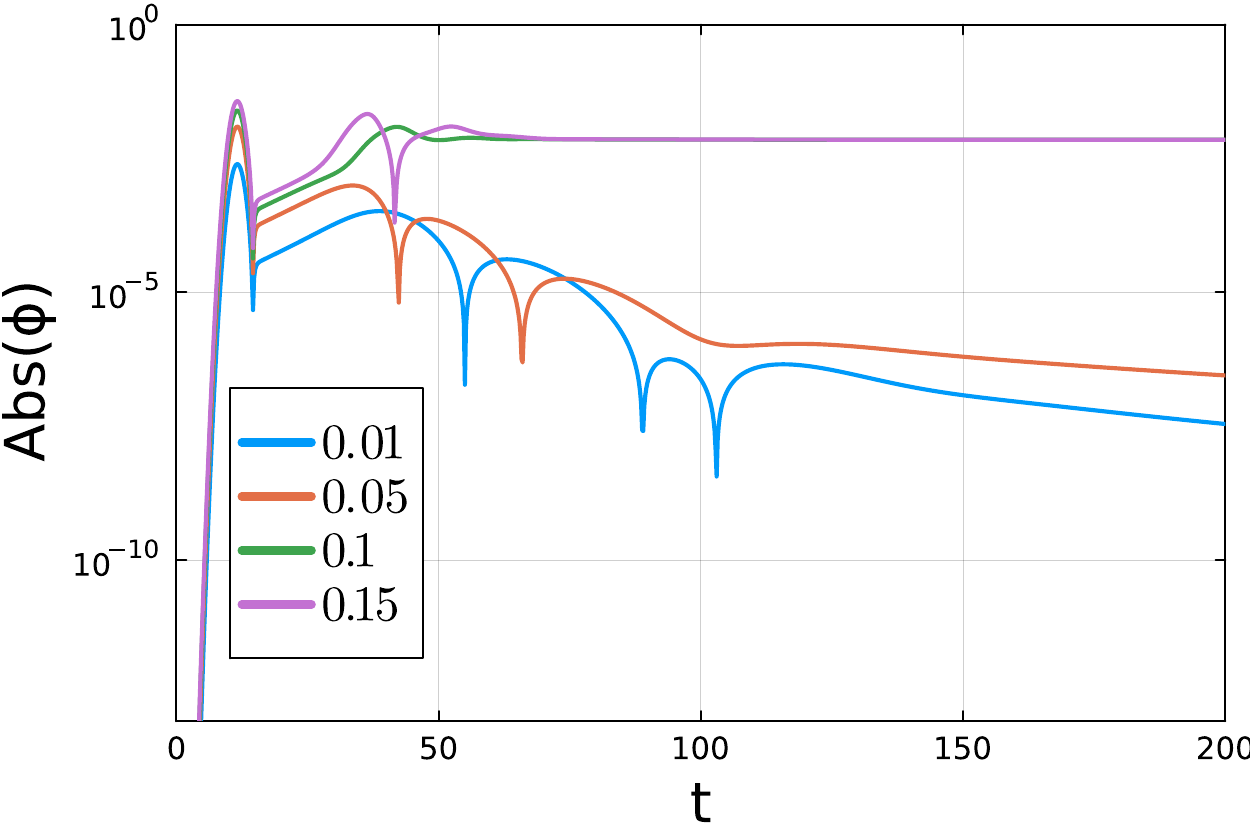}}%
\hfill%
\subfigure[$\zeta_3(\phi)=\alpha\phi^4$]
{\label{time3}
\includegraphics[width=0.3\textwidth]{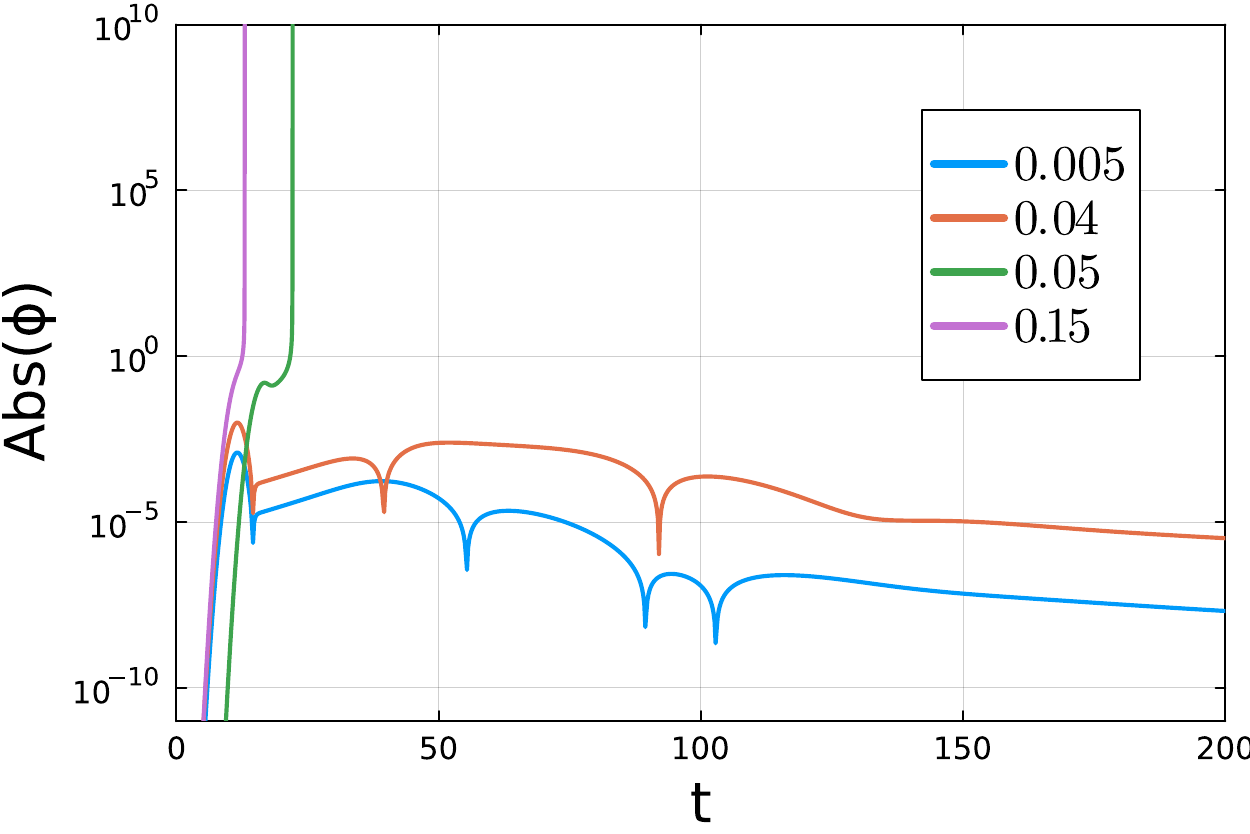}}
\caption{Time evolution of the scalar field on the SBH  background with $M/\lambda=0.025$ for $\alpha=\frac{1}{4}$ and $\beta=\frac{100}{8}$. 
The threshold amplitudes $A_{\rm th}$ are 0.05 (a) and 0.1 (b), indicating the formation of a scalarized phase for $A \ge A_{\rm th}$ for coupling functions $\zeta_1(\phi)$ and $\zeta_2(\phi)$. 
(c) For $\zeta_3(\phi)$ the scalar field diverges within a short timescale for large initial amplitudes or decays for small initial amplitudes of the Gaussian wave.}
\label{fig1}
\end{figure}

\begin{figure}[H]
\centering
\subfigure[$\beta=25/8$]{
\label{time1a} 
\includegraphics[width=0.3\textwidth]{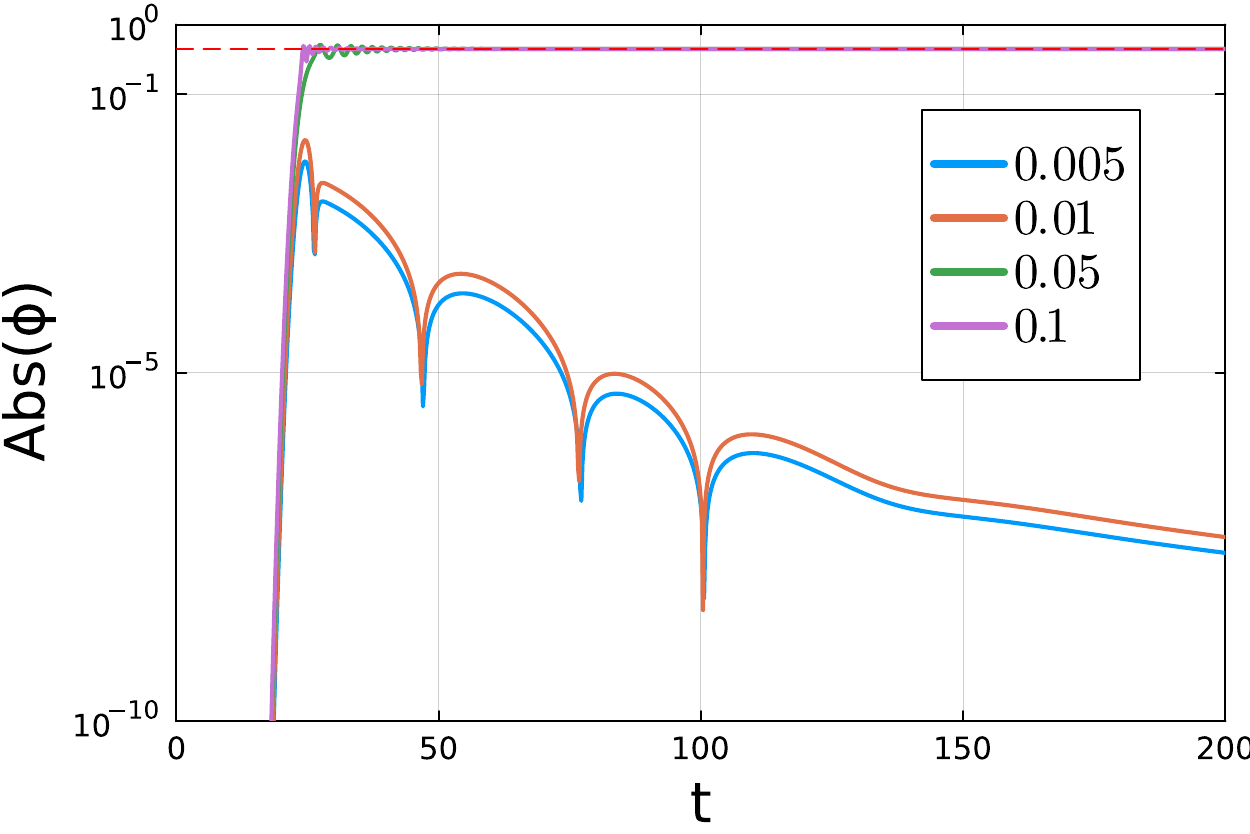}}
\hfill%
\subfigure[$\beta=100/8$]{
\label{time2a} 
\includegraphics[width=0.3\textwidth]{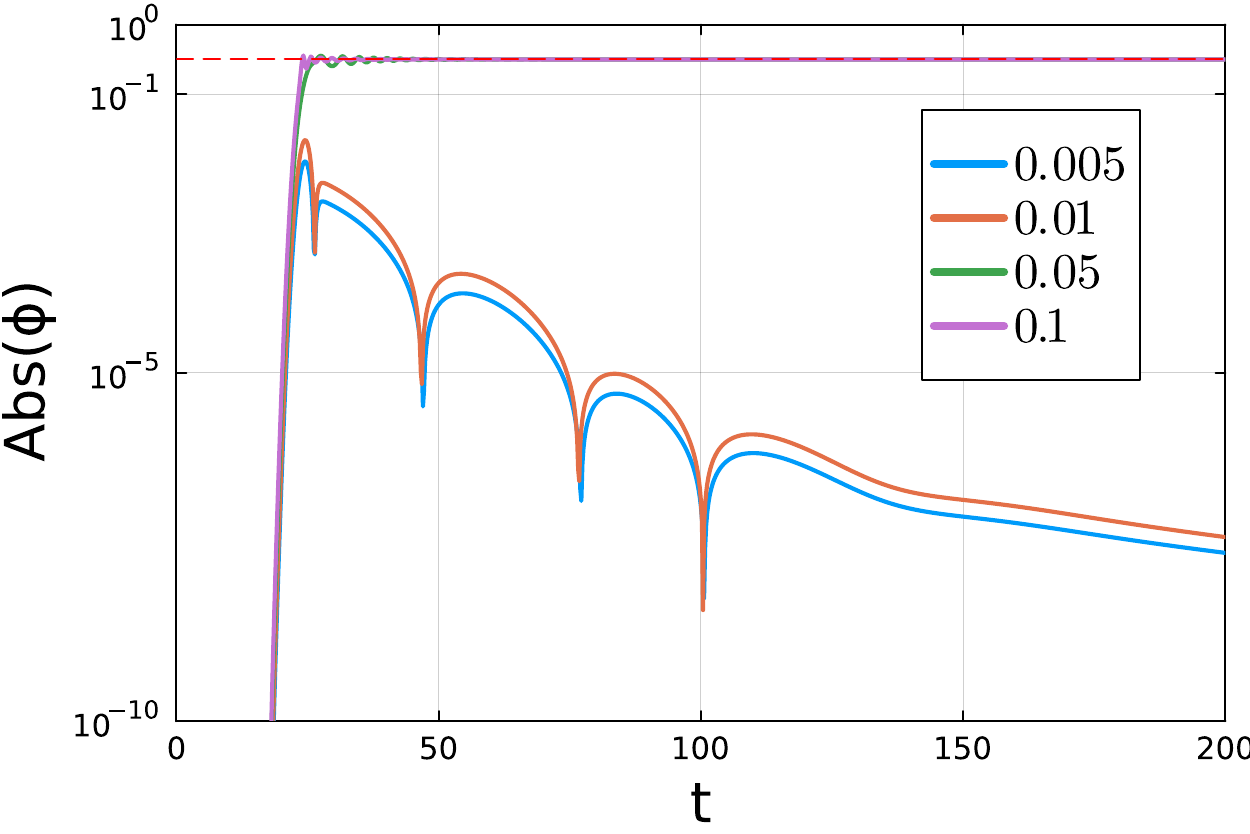}}%
\hfill%
\subfigure[$\beta=1000/8$]{\label{time3a}
\includegraphics[width=0.3\textwidth]{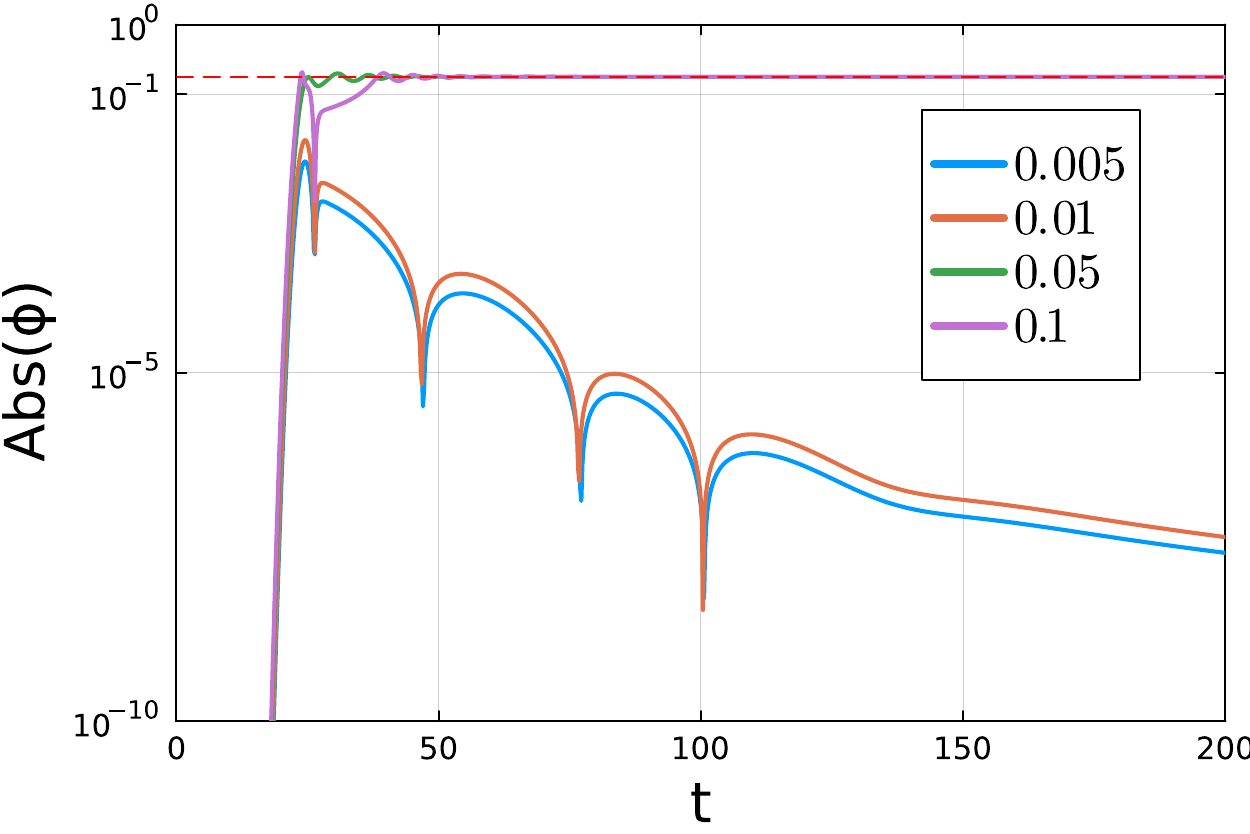}}
\caption{Time evolution of the scalar field on the SBH background with $M/\lambda=0.025$ for $\zeta_1=\frac{1}{4}\phi^4-\beta\phi^8$. 
The horizontal red dashed lines denote the scalar field values $\phi_H$ at the horizon of the final 
configurations: (a) $0.4472$, (b) $0.3162$, (c) $0.1778$. 
}\label{fig1s}
\end{figure}

We exhibit the results of the time evolution in Figs.~\ref{fig1} and \ref{fig1s}.
In Fig.~\ref{fig1} we compare the evolution for the three coupling functions $\zeta_1(\phi)$, $\zeta_2(\phi)$ and $\zeta_3(\phi)$, and locate the observer at $r_\text{ob}=10 r_H$.
In Fig.~\ref{fig1s} we compare the evolution for three values of $\beta$ in the coupling function $\zeta_1(\phi)$, and place a virtual
observer at $r_\text{ob}=1.001 r_H$.
We then slowly increase the amplitude $A$ of the Gaussian wave.
For small amplitudes $A$ the standard exponential decay is observed in Fig.\ref{fig1}. 
But when the amplitude exceeds a certain threshold value which corresponds to $A_\text{th}=0.05$ for coupling function $\zeta_1(\phi)$ (Fig.~\ref{time1}) and to $0.1$ for $\zeta_2(\phi)$ (Fig.~\ref{time2}), the scalar field first grows due to the energy carried by the initial perturbation and then stabilizes to a time-independent configuration that indicates the formation of a scalarized phase.

With regard to the quartic coupling function $\zeta_3(\phi)=\alpha\phi^4$, Fig.~\ref{time3} demonstrates the time evolution of the scalar field perturbations.
In this case, the scalar field $\phi(r_\text{ob},t)$ 
also decays for small initial amplitudes of the Gaussian wave, but it diverges within a short timescale for large initial amplitudes.
This indicates the existence of a runaway instability, and thus no stable scalarized black holes exist. 
In fact, Doneva et al.~\cite{Doneva:2021tvn} pointed out that when the coupling functions are directly proportional to $\phi^4$ and $\phi^6$, there are no stable nonlinearly scalarized solutions similar to the case of tachyonic scalarization in EsGB gravity.

\subsection{Effective potential} 

In order to understand more intuitively the dependence of the scalar field $\phi$ on the coupling function $\zeta_i$, we 
rewrite the Klein-Gordon equation \eqref{KG1} as
\begin{eqnarray}\label{KG2a}
\Box \phi-\partial_{\phi}V^i_\text{eff}=0,
\end{eqnarray}
where we introduce the effective scalar potential 
\begin{eqnarray}\label{Veff}
V^i_\text{eff}=-\frac{\lambda^2}{4}R_{GB}^2\zeta_i(\phi).
\end{eqnarray}
Then, the nonlinear evolutionary states of the scalar can be understood through this effective potential $V^i_\text{eff}$.

\begin{figure}[H]
\centering
\subfigure[Effective potential $V_\text{eff}^3(r,\phi)$]{
\label{figv1} 
\includegraphics[width=0.45\textwidth]{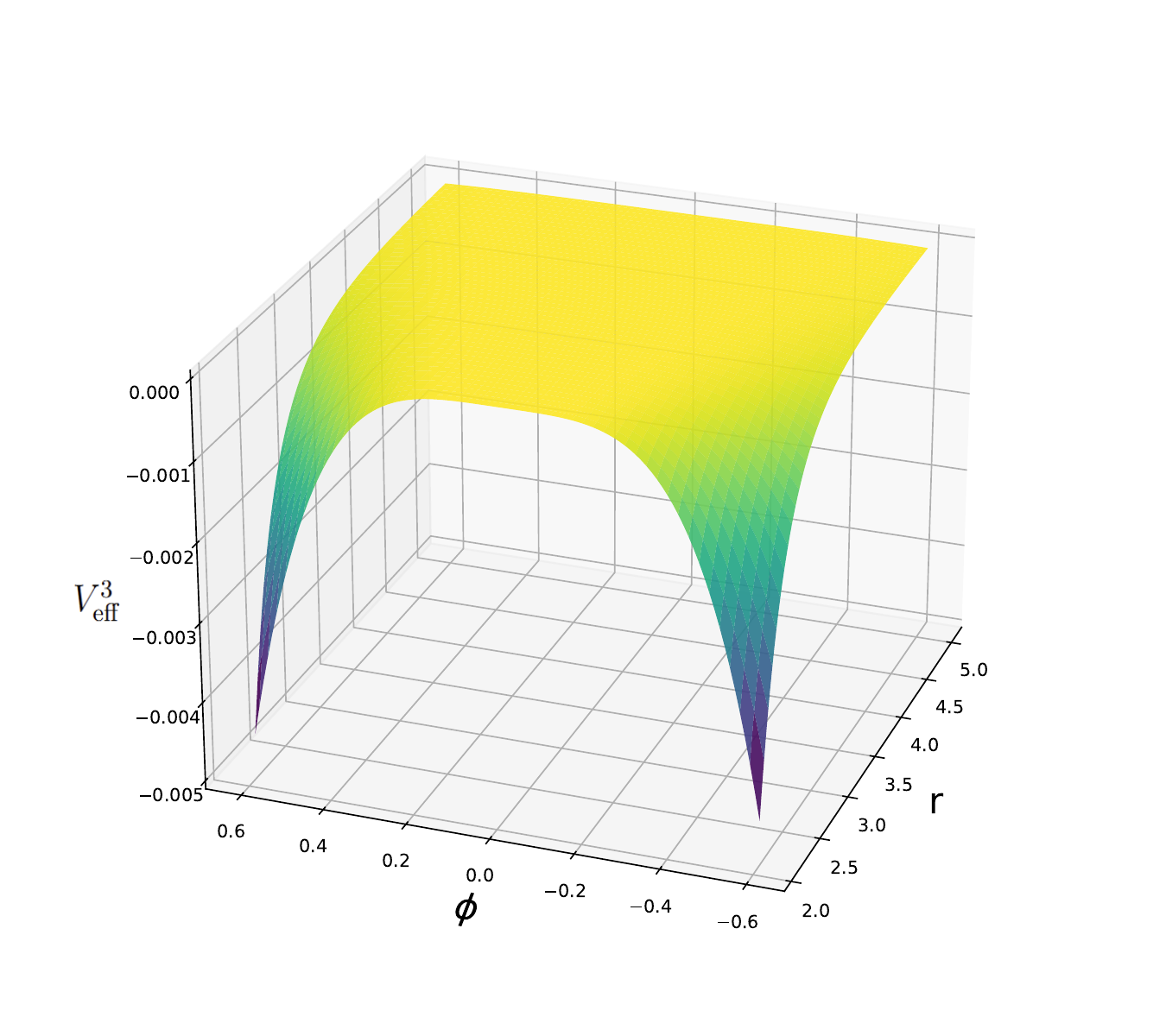}}
\hfill%
\subfigure[Effective potential cut at $r=2.1M$]{
\label{figv2} 
\includegraphics[width=0.45\textwidth]{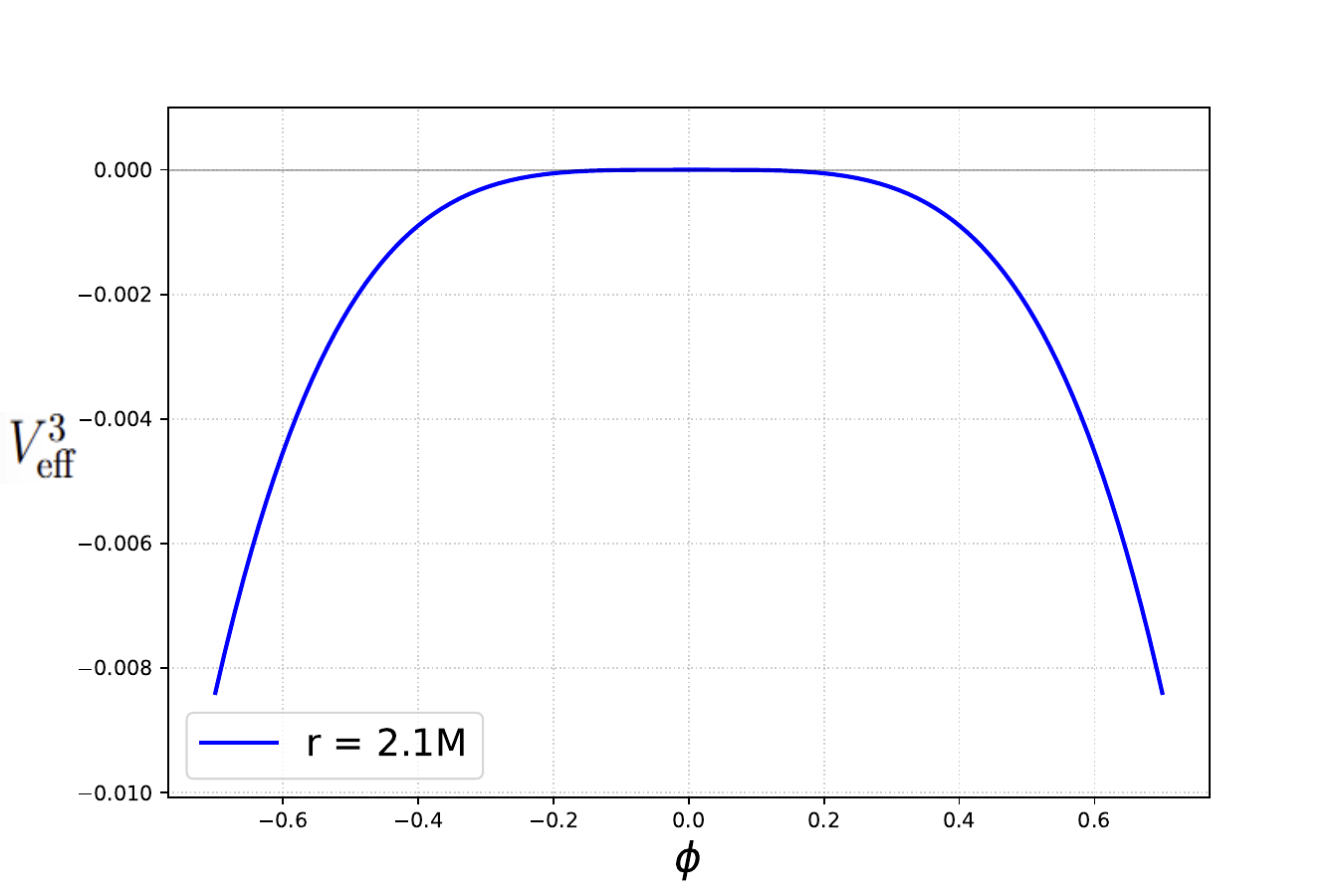}}%
\caption{The effective potential $V^3_\text{eff}$ for $\zeta_3(\phi)=\frac{1}{4}\phi^4$ for $M=1$ and $\lambda=1$. 
(a) Three-dimensional plot of $V^3_\text{eff}(r,\phi)$ as a function of the scalar $\phi$ and the radial coordinate $r$. 
(b) Two-dimensional plot of $V^3_\text{eff}(\phi)$ as a function of $\phi$ for a fixed $r$.}\label{fig1V}
\end{figure}

For the $\zeta_3(\phi)$ coupling, the effective potential $V^3_\text{eff}$ is unbounded from below in the $\phi$-direction
\begin{eqnarray}
V^3_\text{eff}(r,\phi)=-\frac{12\lambda^2 M^2 \alpha}{r^6} \phi^4.
\end{eqnarray}
The trivial SBH solution ($\phi=0$) resides at the local maximum of the potential (see Fig~\ref{fig1V}). 
The scalar profile corresponds to an inverted bowl whose potential energy drops indefinitely as $\phi$ increases.
Then, the scalar undergoes a ``nonlinear runaway'', cascading down the infinite slope. 
This is a signal for the divergence observed in the time evolutions (Fig.~\ref{time3}), showing that the $\zeta_3(\phi)$ coupling is unable to support stable scalar hair.

In contrast, the coupling function $\zeta_1(\phi)$ (and analogously $\zeta_2(\phi)$) may lead to a saturation mechanism in the time evolution. 
For $\zeta_1(\phi)$ the effective potential takes the form
\begin{eqnarray}
V^1_\text{eff}(r,\phi)=-\frac{12\lambda^2 M^2}{r^6} \left(\alpha\phi^4 - \beta\phi^8\right),
\end{eqnarray}
implying the existence of a 
minimum with respect to $\phi$ at $\phi_{\rm s}$.
\begin{eqnarray}
\phi_{\rm s}=\pm\left(\frac{\alpha}{2\beta}\right)^{1/4},
\label{phis}
\end{eqnarray}
The coupling function then satisfies condition \eqref{const} not only at $\phi=0$ but also at $\phi_{\rm s}$. 

\begin{figure}[H]
\centering
\includegraphics[width=0.45\textwidth]{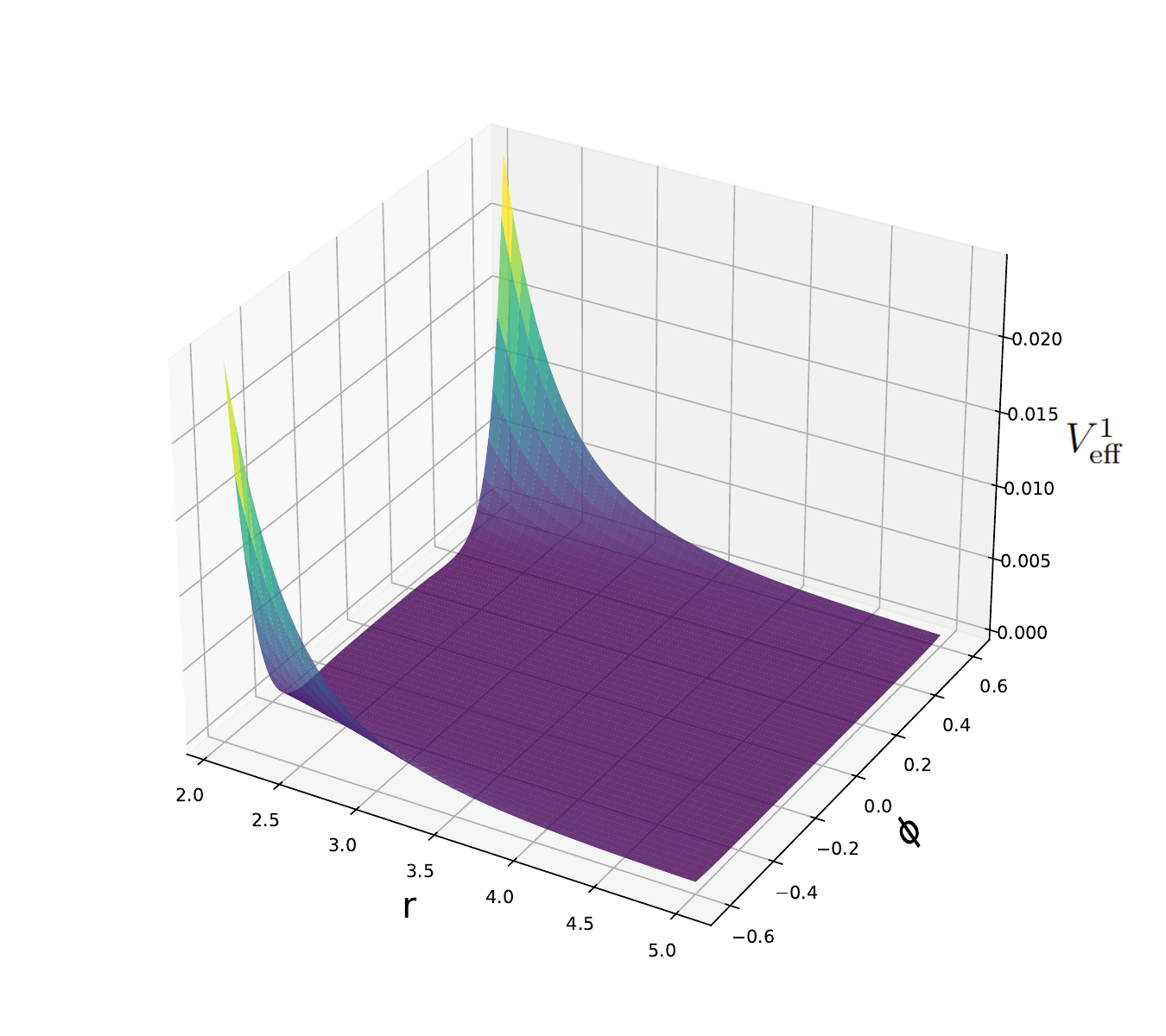}
\caption{Three-dimensional plot of the effective potential $V^1_\text{eff}$ for $\zeta_1(\phi)=\frac{1}{4}\phi^4-\frac{25}{8}\phi^8$ as a function of the scalar $\phi$ and the radial coordinate $r$ for $M=1$ and $\lambda=1$.}  
\label{fig2V}
\end{figure}

\begin{figure}[H]
\centering
\subfigure[Effective potential $V^1_\text{eff}(r,\phi)$]{
\label{figv4} 
\includegraphics[width=0.45\textwidth]{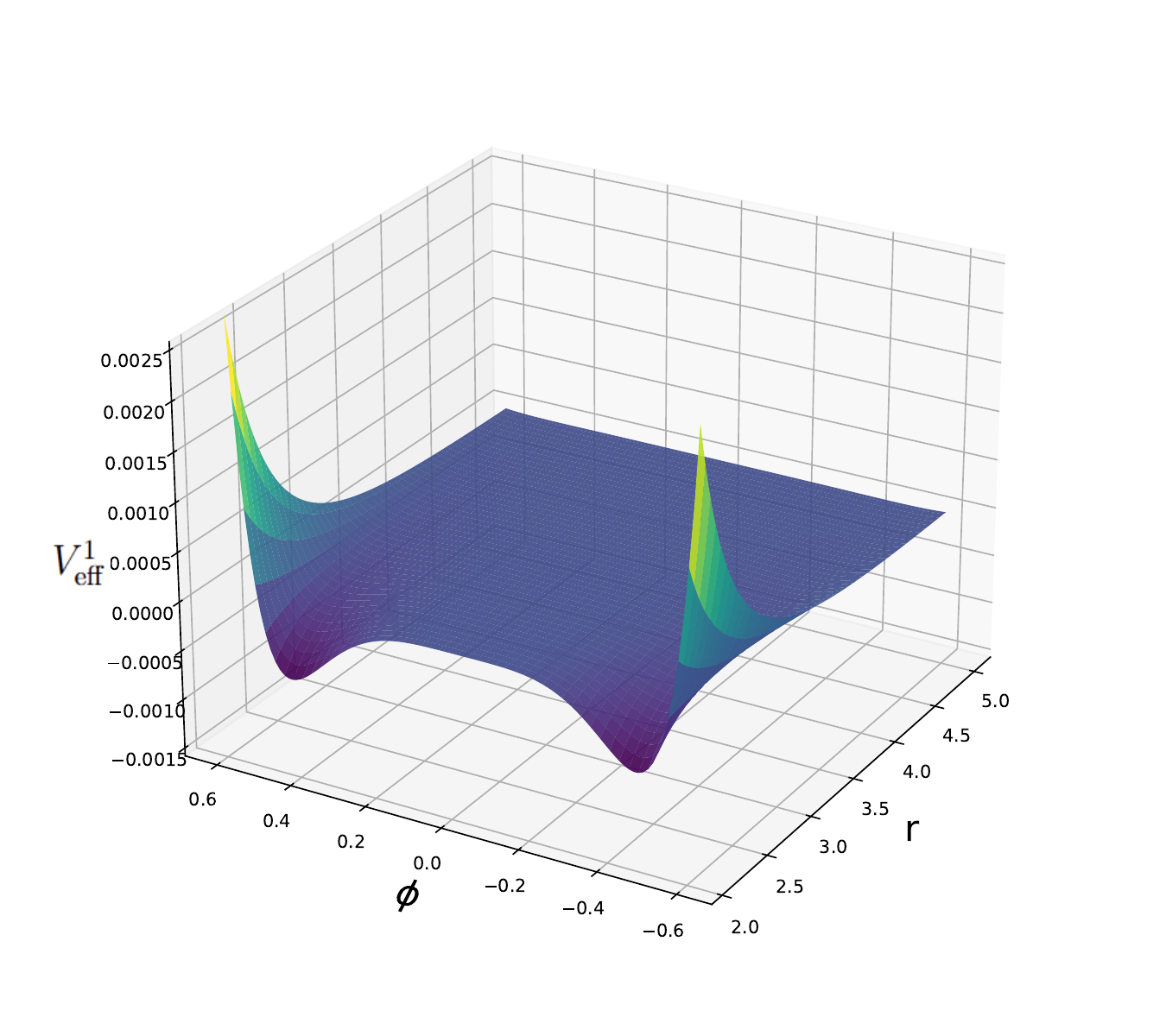}}%
\hfill%
\subfigure[$W$-shaped potential at $r=2.1M$]{\label{figv5}
\includegraphics[width=0.45\textwidth]{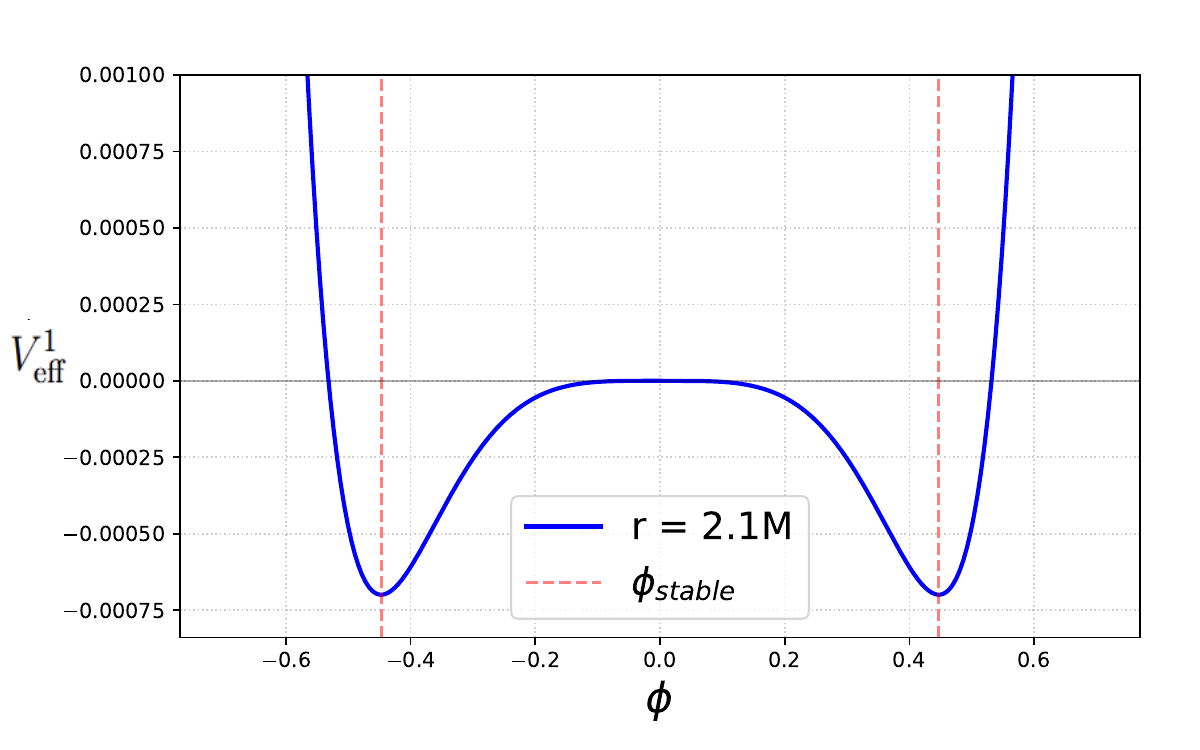}}
\caption{Zoomed figure for the effective potential $V^1_\text{eff}$ for $M=1$ and $\lambda=1$. 
(a) Three-dimensional $W$-shape. 
(b) Two-dimensional $W$-profile of $V^1_\text{eff}(\phi)$ as a function of $\phi$ at a fixed $r$, with the minimum of the scalar 
field located at $\phi_{\rm s}=\pm 0.4472$.}\label{fig2V2}
\end{figure}

Choosing $\zeta_1(\phi)=\frac{1}{4}\phi^4-\frac{25}{8}\phi^8$, a three-dimensional plot of the effective potential $V^1_\text{eff}(r,\phi)$ is shown as a function of $\phi$ and $r$ in Fig~\ref{fig2V}.  
Due to the presence of the $\beta\phi^8$ term, the potential exhibits a steep rise in the near-horizon region, and for larger positive $\phi$, it forms a sharp peak.  
Transitioning into a deep valley elsewhere, it shows an asymmetric structure reminiscent of a ``funnel with steep walls''.  
In Fig.~\ref{fig2V2}, the zoomed figure shows the $W$-shaped potential well clearly.
For large $\phi$, the scalar surmounts the central barrier and is subsequently ``trapped" within this potential well, where the positive potential well is located at $\phi_{\rm s}=0.4472$. 
Consequently, this system settles into a stationary configuration, as shown in the plateau of Fig.~\ref{time1}. 
This implies that the $\beta\phi^8$ term provides a repulsive feedback that balances the attractive term ($\alpha\phi^4$) at large $\phi$. 
This is essential for forming stable scalarized black holes.

We finally present the (positive) minima of the scalar field for a set of coupling constants for later comparison 
\begin{eqnarray}\label{phistable}
\phi_{\text{s}} =
\begin{cases}
\frac{1}{\sqrt{5}}=0.4472 & \text{for }  \alpha=1/4, \quad  \beta=25/8 \quad (i),\\
\frac{1}{\sqrt{10}}=0.3162 & \text{for } \alpha=1/4, \quad \beta =100/8 \quad (ii), \\
\frac{1}{10^{3/4}}=0.1778 & \text{for } \alpha=1/4, \quad  \beta = 1000/8 \quad (iii).
\end{cases}
\end{eqnarray}

\section{Numerical solutions of scalarized black holes}
\label{Sec3}

\subsection{Theoretical setup}

After gaining insight when scalarized phases may exist and form dynamically, we now study the scalarized black holes by solving the nonlinear system of coupled field equations.
For the spherically symmetric spacetimes we employ the metric ansatz
\begin{eqnarray}
d{s^2} \equiv g_{\mu\nu}dx^\mu dx^\nu= -A (r)d{t^2} + \frac{d{r^2}}{B(r)} + {r^2}d\Omega^2_2.
\label{metric}
\end{eqnarray}
Substituting the above ansatz into the field equations \eqref{field1} and \eqref{KG1}, we obtain the gravitational field equations
\begin{eqnarray}\label{graveq}
E_{tt}&=&1-rB'-B\left(1+r^2\phi'^2\right)+2(3B-1)B'{\psi _r}+4B(B-1){\psi _r}' = 0,\\
E_{rr}&=&\Big[2\left(1-3 B\right) {\psi _r} +r\Big] BA'+AB\left(1-r^2\phi '^2\right)-A= 0,\\
E_{\theta\theta}&=&4B\Big[\psi_r A' \left(B A'-3 A B'\right)-2 A B \left(A' \psi_r\right)'\Big]+4 r
A^2 B\phi'^2\\
&&+A\Big[2\left(B A\right)'+r
\left(A' B\right)'+r A B\left(\frac{A'}{A}\right)'\Big],
\end{eqnarray}
while the scalar field equation is given by
\begin{eqnarray}\label{KG3}
&&\phi ''+ \frac{1}{2} \phi '
   \left(\frac{A'}{A}+\frac{B'}{B}+\frac{4}{r}\right)\nonumber\\
 && \quad \quad +\lambda^2\left(\frac{(B-1) A''}{r^2 A}+\frac{(3 B-1)
   A' B'}{2 r^2 A B}-\frac{(B-1) A'^2}{2 r^2
   A^2}\right)\frac{d\zeta(\phi)}{d\phi} =0
\end{eqnarray}
with
\begin{eqnarray}
{\psi _r} = {\lambda ^2}\frac{d\zeta(\phi)}{d\phi}\frac{{d\phi }}{{dr}}.
\end{eqnarray}

In order to obtain asymptotically flat black hole solutions that are regular on and outside the horizon and carry scalar hair, boundary conditions near the horizon are required,
\begin{eqnarray}
&A(r)\left| {_{r \to r_H} \to 0,} \quad\right.B(r)\left| {_{r \to r_H} \to 0, }\right. 
\end{eqnarray}
and at asymptotic infinity
\begin{eqnarray}
&A(r) \left| {_{r \to \infty } \to 1,\quad B(r)\left| {_{r \to \infty }} \right.} \right. \to 1,\quad\phi(r) \left| {_{r \to \infty }} \right. \to 0.
\end{eqnarray}
The asymptotic behavior for $r \to \infty$ determines also the global charges, the mass $M$ and the scalar charge $Q_s$
\begin{equation}
    A(r) = 1 - \frac{2M}{r} + \cdots , \quad 
    \phi(r) = \frac{Q_s}{r} + \cdots \ .
    \label{charges}
\end{equation}

Moreover, the regularity of the first derivative of the scalar field at the horizon imposes the condition \cite{Doneva:2017bvd} for the existence of scalarized black holes
\begin{eqnarray}
\label{condition}
\Delta \equiv1-\frac{24\lambda^4}{r_H^4}\left(\frac{d\zeta(\phi)}{d\phi}(\phi_H)\right)^2 > 0.
\end{eqnarray}
For $\Delta=0$ we obtain the critical curve, which delimits the domain of existence of nonlinearly scalarized black holes for a given coupling function
\begin{equation}
    r_H = \left(24\lambda^4 \left(\frac{d\zeta_1(\phi)}{d\phi}(\phi_H)\right)^2 \right)^{1/4} .
    \label{cond2}
    \end{equation}
Thus, for the constant scalar field solutions $\phi=0$ and $\phi=\phi_{\rm s}$ the values of the horizon radius on the critical curve are $r_H=0$.

Finally, the Kretschmann scalar is given by
\begin{eqnarray}
R^2_K&\equiv& R_{\mu\nu\rho\sigma}R^{\mu\nu\rho\sigma}=\frac{B'^2}{4}\left(\frac{A'^2}{A^2}+\frac{8}{r^2}\right)
  +\frac{B^2 A'^2 \left(r^2 A'^2+8 A^2\right)}{4
   r^2 A^4}+\frac{B A''\left(B A'\right)'}{A^2}\nonumber\\
   &&-\frac{B A'\left(A'^2 B'\right)'}{2 A^3}+\frac{4(B-1)^2}{r^4}.\label{K}
\end{eqnarray}

\subsection{Results for the polynomial coupling function $\zeta_3(\phi)$}



We start the discussion of the nonlinearly scalarized solutions with coupling function $\zeta_3(\phi)$ (fixing $\alpha=1/4$), which corresponds the taking $\beta=0$ in $\zeta_1(\phi)$ and $\zeta_2(\phi)$.
In this case, there is only a single solution with a constant scalar field, $\phi=0$.
The time evolution method did not lead to stationary solutions.
Solving, however, the time-independent scalar equation in the SBH background does yield solutions.
They are referred to as the probe limit or decoupling limit.

\begin{figure}[H]
\centering
\subfigure[$\phi_H$ vs $ M$]{
\label{3ax} 
\includegraphics[width=0.3\textwidth]{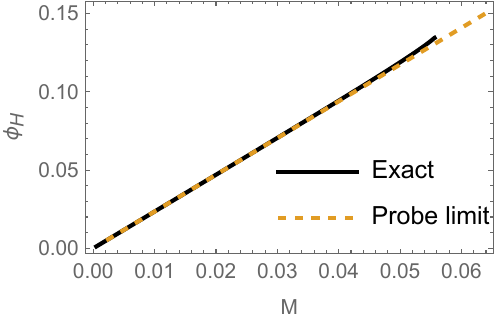}}
\hfill
\subfigure[$Q_s$ vs $ M$]{
\label{3bx} 
\includegraphics[width=0.3\textwidth]{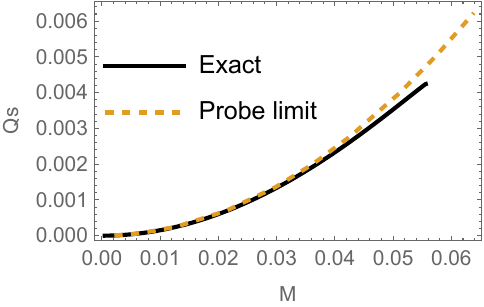}}
\hfill
\subfigure[$\Delta$ vs $ M$]{
\label{3cx} 
\includegraphics[width=0.3\textwidth]{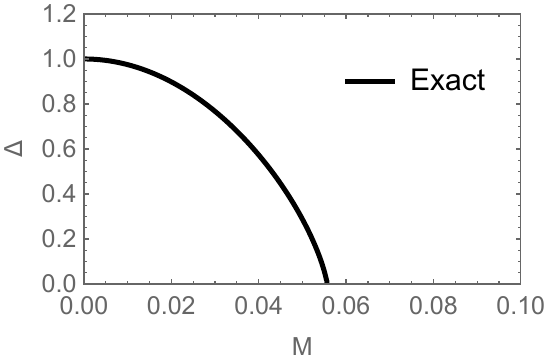}}
\caption{Scalar field $\phi_H$ at the horizon, scalar charge $Q_s$, and regularity parameter $\Delta$ as functions of the mass $M$ for the coupling function $\zeta_1(\phi)$ for $\alpha=1/4$, $\beta=0$, $\lambda=1$.} 
\label{fig3a}
\end{figure}

To solve the scalar field equation, we define the field as 
$\phi(r)=\frac{r_H}{\lambda} \psi(\frac{r}{r_H})$.
Substituting this ansatz into the scalar field equation yields a dimensionless differential equation for $\psi(\frac{r}{r_H})$ that is independent of any physical parameters.
Consequently, the solution depends solely on the dimensionless radius, implying that $\psi(1)=c_0$ is a universal constant.
Evaluating the field at the horizon gives $\phi(r_H)= \frac{r_H}{\lambda} c_0$.
Thus, $\frac{\phi_H}{r_H}=\frac{c_0}{\lambda}=1.17443$.
We therefore obtain a linear dependence for the scalar field $\phi_H$ at the horizon and the mass $M$ in the probe limit
\begin{equation} 
\phi_H=2.34885 M,
\label{probe}
\end{equation}
as shown in Fig.~\ref{fig3a}. 
The corresponding dependence of the scalar charge $Q_s$ on the mass $M$ in the probe limit is seen in Fig.~\ref{3bx}.
We note that the regularity condition \label{condition} does not arise in the probe limit.

Since the the coupling function has a single extremum with respect to $\phi$, $\phi=0$, we obtain only a single branch of scalarized solutions, which originates at $\phi_H=0=M$.
The solutions of the coupled set of field equations where backreaction is taken into account are also exhibited in Fig.~\ref{fig3a}.
Here we show the scalar field $\phi_H$ at the horizon in Fig.~\ref{3ax},
the scalar charge $Q_s$ in Fig.~\ref{3bx}, and the regularity parameter $\Delta$ in Fig.~\ref{3cx} versus the mass $M$.
All three curves are monotonic.

Along this branch, the scalar field remains small.
Thus, there is only little backreaction of the scalar field on the metric, which remains rather close to the Schwarzschild metric.
The branch ends when the regularity parameter $\Delta$ reaches the singular limit $\Delta=0$. 
A reasonably good estimate for this endpoint is obtained by determining the crossing of the line \eqref{probe} with the curve \eqref{cond2} using the SBH relation $r_H=2M$.
As we will see in the following, this branch is rather universal.
However, it is unstable \cite{Blazquez-Salcedo:2022omw}.

\subsection{Results for the polynomial coupling function $\zeta_1(\phi)$}

We now consider the coupling function $\zeta_1(\phi)$ for three values of $\beta$
(fixing $\alpha=1/4$)~\footnotemark[1]
\footnotetext[1]{We choose $\alpha=1/4$ to be compatible with the expansion of the exponential coupling function in Refs.~\cite{Doneva:2021tvn,Blazquez-Salcedo:2022omw}.}.
In the probe limit we obtain the same pattern, independent of the choice of $\beta >0$.
We always have two branches of solutions, since there is in addition to $\phi=0$ the constant scalar field solution $\phi=\phi_{\rm s}$.
In a $\phi_H$-$M$ diagram the linear solution for $\beta=0$ \eqref{probe} and the solution constant solution $\phi_{\rm s}$ form a triangle which delimits the domain of existence of scalarized black holes.

Already in the probe limit with a finite value of $\beta$, however, the effects of this nonlinear higher order term become visible by causing deviations from the linear $\beta=0$ case, which become stronger as this nonlinear term increases.
The solutions then reach a maximal value of the mass $M$ shortly before $\phi_H$ approaches $\phi_{\rm s}$.
Here it merges with the upper branch, the primary branch, that starts at $M=0$, and features an almost constant value of $\phi_H \approx \phi_{\rm s}$ except close to the merger point.
This upper branch is the primary branch, that is found in the time-dependent calculations.
We note, that in the numerical calculations, we lose accuracy on this upper branch when approaching $M=0$.
Therefore these probe limit curves in Figs.~\ref{fig3}, \ref{fig5} and \ref{fig6} do not extend back to $M=0$.


\subsubsection{ $\beta=25/8$}

\begin{figure}[H]
\centering
\subfigure[$\phi_H$ vs $ M$]{
\label{3a} 
\includegraphics[width=0.3\textwidth]
{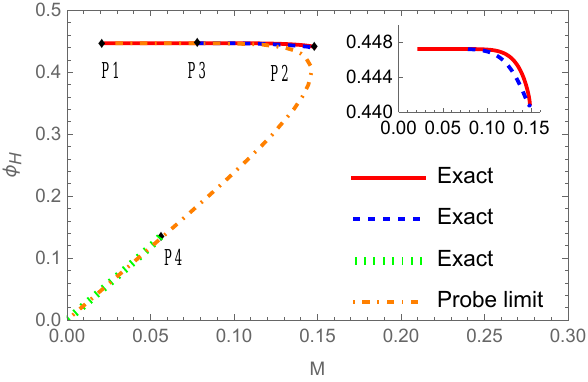}
}
\hfill
\subfigure[$Q_s$ vs $ M$]{
\label{3b} 
\includegraphics[width=0.3\textwidth]
{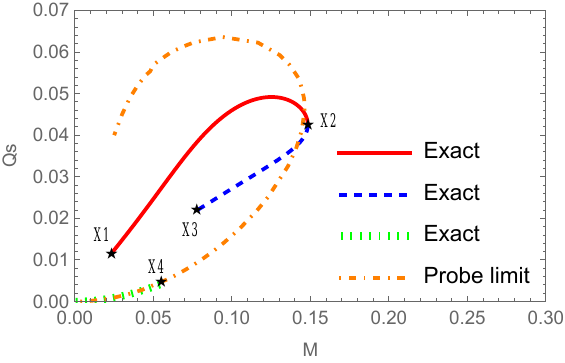}
}
\hfill
\subfigure[$\Delta$ vs $ M$]{
\label{3c} 
\includegraphics[width=0.29\textwidth]{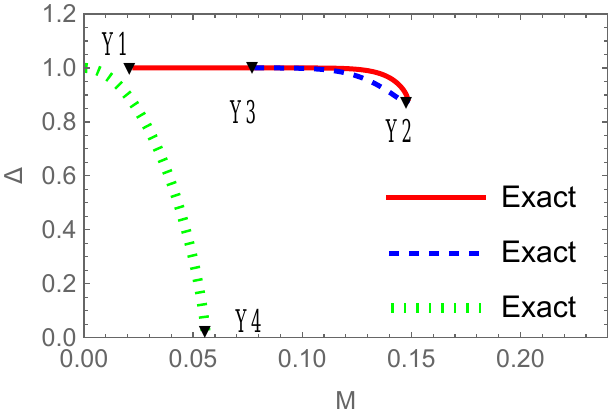}}
\caption{Scalar field $\phi_H$ at the horizon, scalar charge $Q_s$, and regularity parameter $\Delta$ as functions of the mass $M$ for the coupling function $\zeta_1(\phi)$ for $\alpha=1/4$, $\beta=25/8$, $\lambda=1$.} 
\label{fig3}
\end{figure}

We now consider the coupling function $\zeta_1(\phi)$ for small values of $\beta$.
In that case, the upper value $\phi_H=\phi_{\rm c}$ is large.
Clearly, the $\beta$-dependent higher order term is expected to lead to significant deviations from the probe limit by causing stronger back reaction effects.
For $\beta=25/8$, we then obtain three branches (solid, dashed, dotted) of scalarized black hole solutions, as seen in Fig.~\ref{fig3}. 
Fig.~\ref{3a} illustrates the dependence of $\phi_H$ on the mass $M$.
We immediately recognize the lower branch (dotted line) emerging from $\phi_H =0 = M$, 
and terminating at the point $P4$ ($M \approx 0.05592$, $\phi_H \approx 0.1353$) in Fig.~\ref{3a}.
Here the regularity parameter $\Delta$ drops to zero
as seen Fig.~\ref{3c}. 
This branch is almost identical to the single branch of the $\zeta_3(\phi)$ coupling function, since the scalar field is still small along this branch.
Moreover, since $\beta$ multiplies a high power of $\phi$, the influence of $\beta$ along this branch is almost negligible.
Since the endpoint of the branch is associated with a value of $\phi_H$ that is considerably smaller than $\phi_{\rm s}$ the other branches must be disconnected from the lower branch.

\begin{figure}[h]
\centering
\subfigure[$R^2_K$ for point `1' ($X1$, $P1$, $Y1$)]{
\label{4a} 
\includegraphics[width=0.32\textwidth]{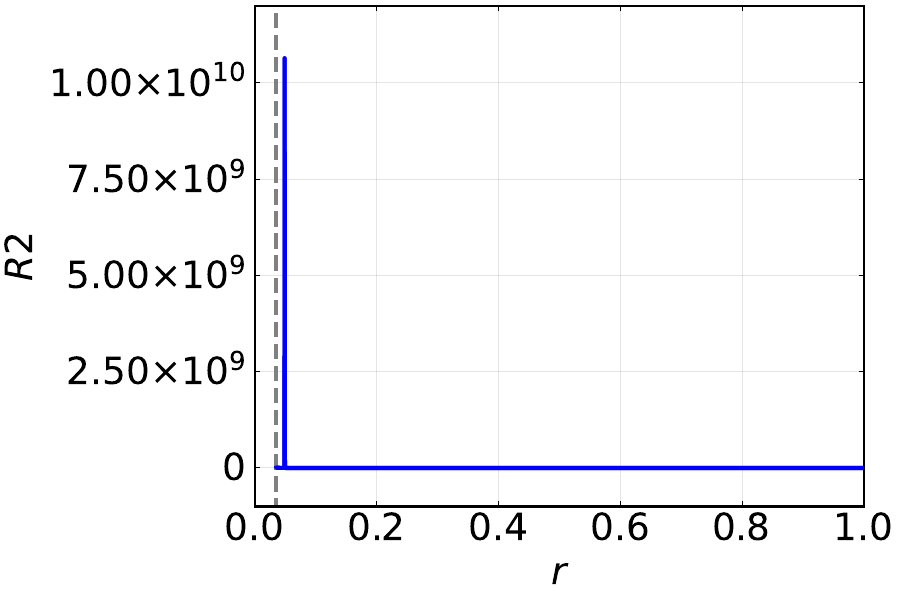}}%
\hspace{0.5cm}
\subfigure[$R^2_K$ for point `3' ( $X3$, $P3$, $Y3$)]{
\label{4b} 
\includegraphics[width=0.32\textwidth]{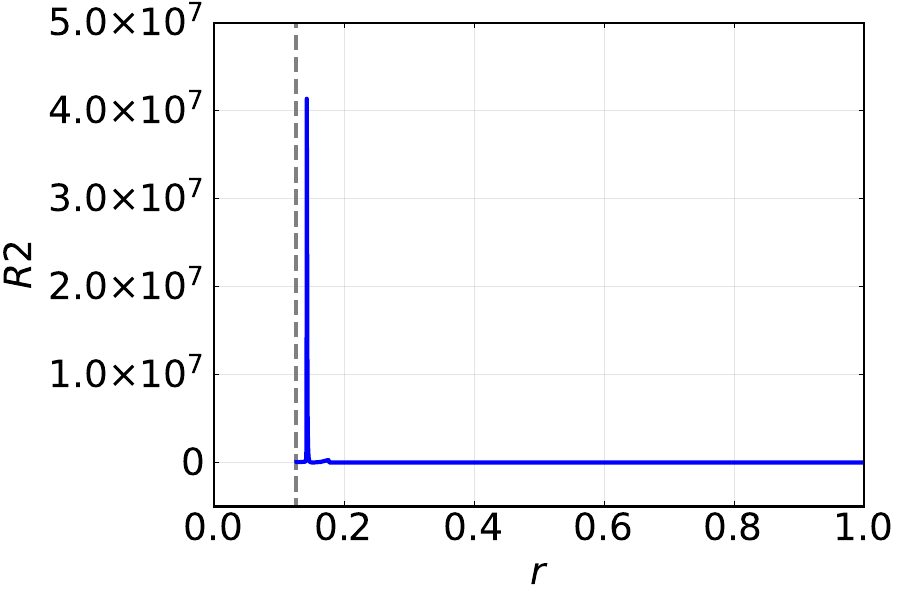}}%
\caption{Divergence of the Kretschmann scalar $R^2_K$ of scalarized black holes with $\zeta_1(\phi)$ and $\beta=25/8$. 
The vertical dashed lines indicate the horizon radius. 
(a): $R^2_K$ for $r_H\approx0.03450$ and $M\approx0.02344$. 
(b): $R^2_K$ for $r_H\approx0.1265$ and $M\approx0.07839$.}
\label{fig4}
\end{figure}

The solid curve in Fig.~\ref{3a} represents the primary branch, that was reached by time evolution and thus may feature stable solutions.
Being associated with $\phi_{\rm s}$, the scalar field along this branch has large horizon values. 
In fact, it starts at $\phi_H \approx 0.4472 = \phi_{\rm s}$ (point $P1$) in the small-mass region with mass $M \approx 0.02344$ and horizon radius $r_H \approx 0.03450$. 
The reason it does not start from $r_H=0=M$ is that the Kretschmann scalar develops a divergence outside the horizon for $r_H < 0.03450$, as seen in Fig.~\ref{4a}, and thus a curvature singularity arises.
Therefore, the mass $M\approx0.02344$ is the minimum mass for scalarized black holes on this branch (Fig.~\ref{3a}).
Along the primary branch, $\phi_H$ decreases only slowly with increasing $M$.
The branch ends at the junction point $P2$ $(M \approx 0.1486, \phi_H \approx 0.4220)$, where it bifurcates with a distinct scalarized branch.

This third branch bends back toward smaller masses (shown by the dashed curve in Fig.~\ref{3a}), 
and terminates at point $P3$ $(M \approx 0.07839,\phi_H\approx0.4472)$. 
Here the Kretschmann scalar develops another divergence outside the horizon, now at $r_H \approx 0.1265$ as seen in Fig.~\ref{4b}. 
Thus, the allowed scalarized black hole solutions are confined to a finite interval of the mass parameter.
A rough estimate of the domain of existence is obtained by considering the crossing of the extended (almost) linear small $\phi$ line with the line $\phi=\phi_{\rm s}$.
The maximal mass ($P2$) is therefore controlled by the value of $\beta$.

Figure \ref{3b} shows the corresponding diagram for the scalar charge $Q_s$ as a function of the mass $M$, where the critical points are denoted $X_1 - X_4$.
Finally, Fig.~\ref{3c} depicts the regularity parameter $\Delta$.
For the points $Y1, Y2$, and $Y3$, $\Delta>0$, indicating that the corresponding solutions satisfy the regularity condition at the horizon.  
On the other hand, at $Y4$ 
the regularity parameter $\Delta$ vanishes.

\subsubsection{$\beta=100/8$}

\begin{figure}[H]
\centering
\subfigure[$\phi_H$ vs $ M$]{
\label{5a} 
\includegraphics[width=0.3\textwidth]{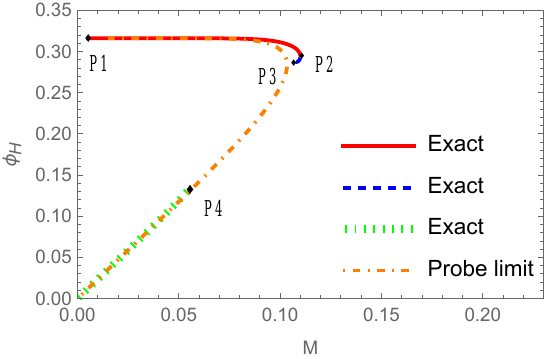}
}
\hfill
\subfigure[$Q_s$ vs $ M$]{
\label{5b} 
\includegraphics[width=0.3\textwidth]{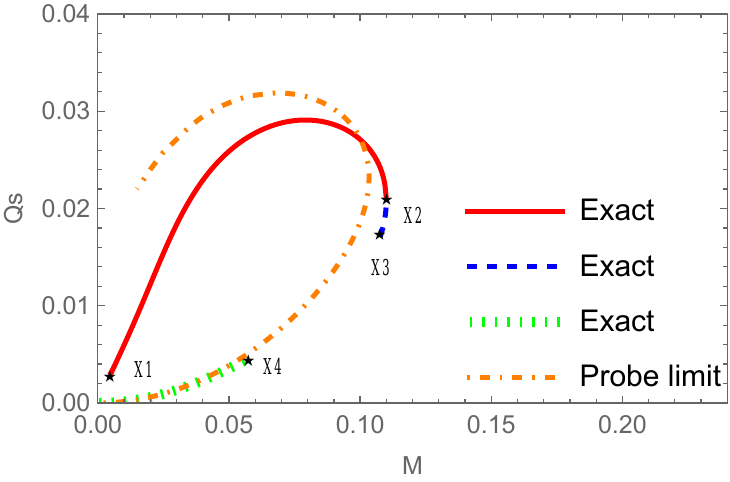}
}
\subfigure[$\Delta$ vs $ M$]{
\label{5c} 
\includegraphics[width=0.29\textwidth]{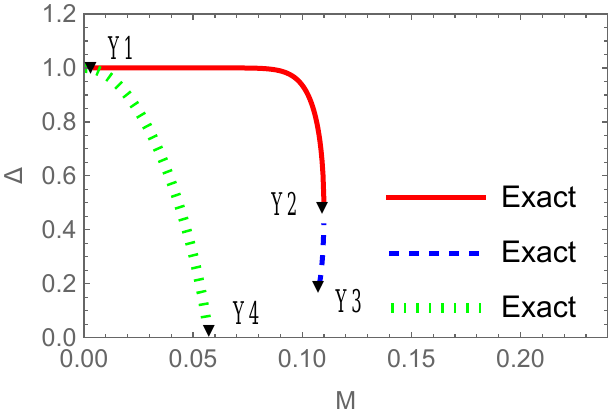}}
\caption{Scalar field $\phi_H$ at the horizon, scalar charge $Q_s$ and regularity parameter $\Delta$ as functions of the mass $M$ for the coupling function $\zeta_1(\phi)$ for $\alpha=1/4$, $\beta=100/8$, $\lambda=1$.}
\label{fig5}
\end{figure}
For intermediate values of $\beta$, e.g., for $\beta=100/8$, the number of branches can still be three, as seen in Fig.~\ref{fig5}. 
As before, we immediately recognize the lower branch (dotted line), starting at $\phi_H =0 =M$ and terminating at $P4$ where $\Delta \to 0$.
Also, the primary branch (solid line) starts at $\phi_H \approx 0.3162=\phi_{\rm s}$ 
and ends at $P2$, where it merges with the third branch (dashed line), which bends backward. 
This branch terminates at $P3$ when the Kretschmann scalar diverges.

\subsubsection{$\beta=1000/8$}

\begin{figure}[H]
\centering
\subfigure[$\phi_H$ vs $ M$]{
\label{6a} 
\includegraphics[width=0.32\textwidth]{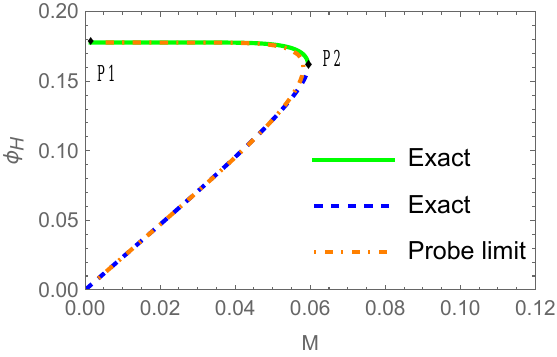}
}
\hfill%
\subfigure[$Q_s$ vs $ M$]{
\label{6b} 
\includegraphics[width=0.32\textwidth]{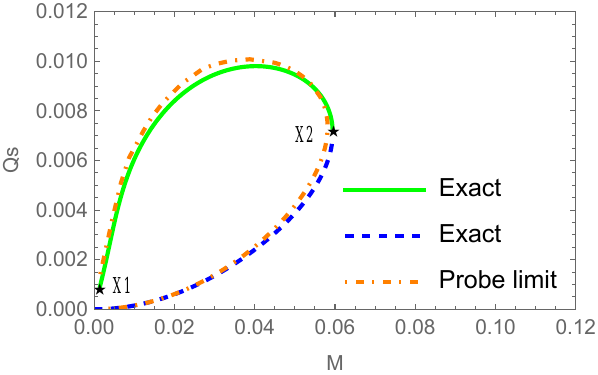}
}
\hfill%
\subfigure[$\Delta$ vs $ M$]{
\label{6c} 
\includegraphics[width=0.3\textwidth]{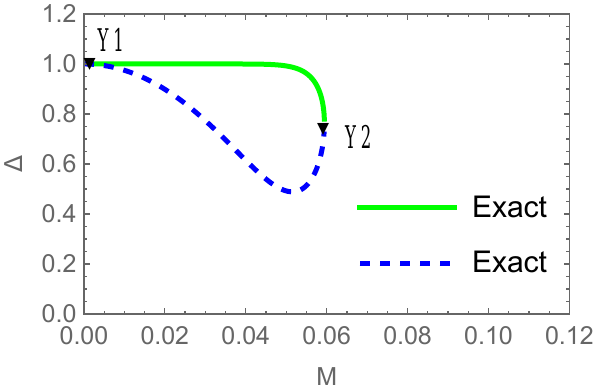}}
\caption{Scalar field $\phi_H$ at the horizon, scalar charge $Q_s$ and regularity parameter $\Delta$ as functions of the mass $M$ for the coupling function $\zeta_1(\phi)$ for $\alpha=1/4$, $\beta=1000/8$, $\lambda=1$. 
}\label{fig6}
\end{figure}

\begin{figure}[H]
\centering
{\includegraphics[width=0.32\textwidth]{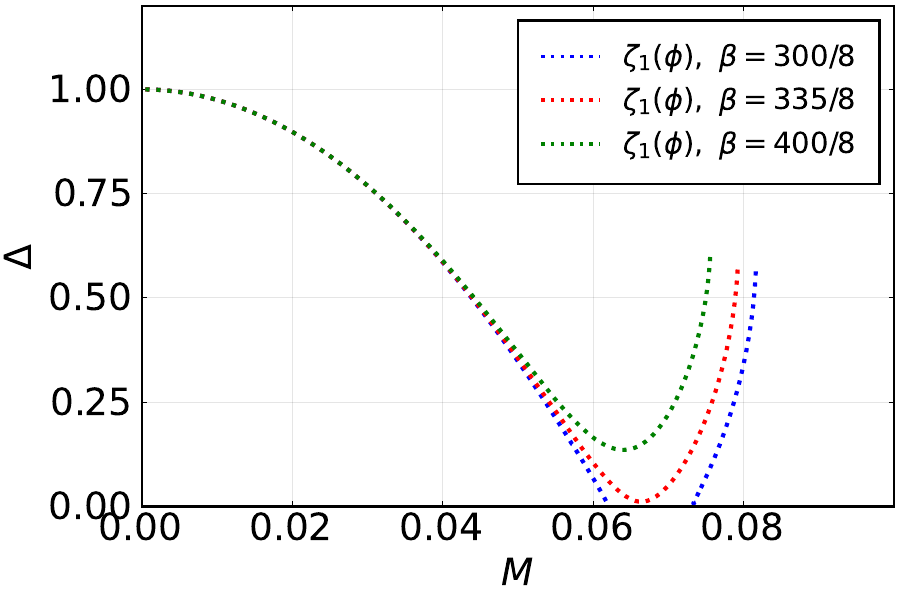}}
\caption{Regularity parameter $\Delta$ as a function of the  mass $M$ for the coupling function $\zeta_1(\phi)$ for $\alpha=1/4$, several values of $\beta$, $\lambda=1$.}\label{fig7}
\end{figure}

For sufficiently large values of $\beta$ such as $\beta=1000/8$, 
the pattern is very similar to the probe limit.
Thus there exist only two branches of scalarized black hole solutions.
The primary branch (solid line) starts from $\phi_H \approx 0.1778=\phi_{\rm s}$ and a small mass at $P1$. 
It extends to $P2$, where it bifurcates with the lower branch (dotted line), as seen in Fig.~\ref{fig6}, that  
starts at $\phi_H=M=0$.
This branch encounters neither a curvature singularity nor $\Delta=0$.
To illustrate the transition from three to two branches, we depict the regularity parameter $\Delta$ for several values of $\beta$ close to the transition in Fig.~\ref{fig7} where the primary branch is omitted for clarity.

\subsubsection{Profile functions}

\begin{figure}[H]
\centering
\subfigure[ $A(r)$, $B(r)$ and $f(r)$ vs $r$]{
\label{8a} 
\includegraphics[width=0.3\textwidth]{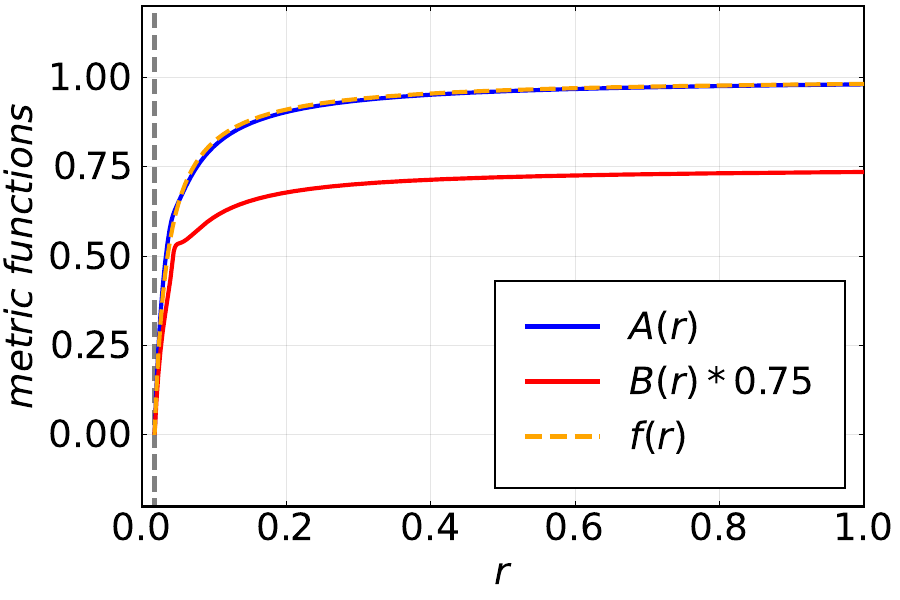}}
\hfill%
\subfigure[ $\delta A(r)$ and $\delta B(r)$ vs $r$]{
\label{8b}
\includegraphics[width=0.3\textwidth]{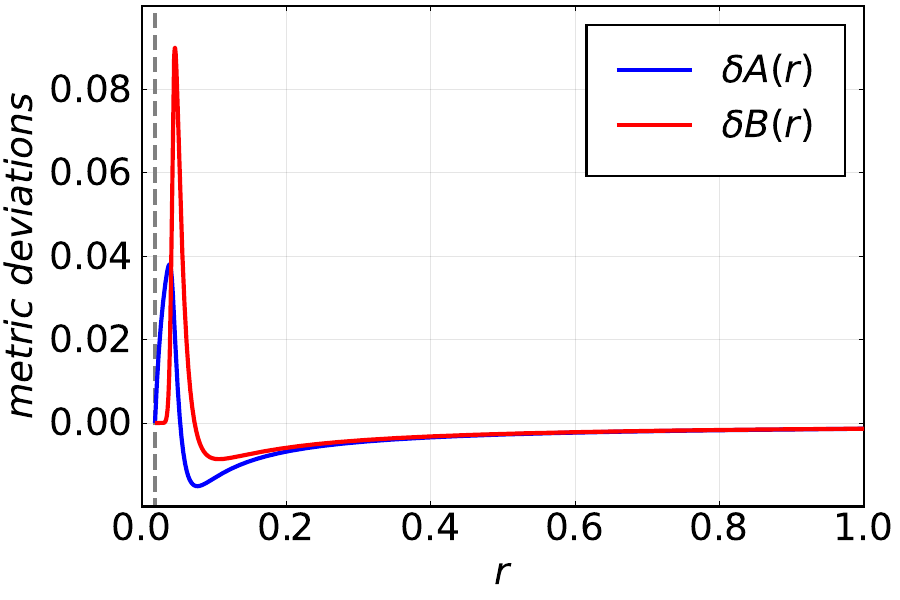}}
\hfill%
\subfigure[$\phi(r)$ vs $r$]{
\label{8c}
\includegraphics[width=0.3\textwidth]{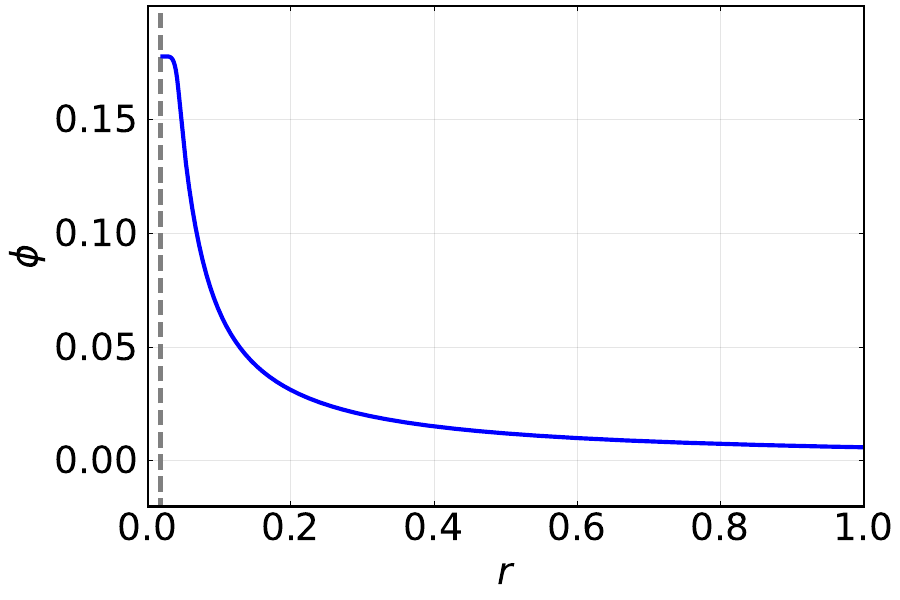}}
\caption{Profile functions of the scalarized black hole on the primary branch and the SBH with $r_H= 0.018$ for the coupling function $\zeta_1(\phi)$ for $\alpha=1/4$, $\beta = 1000/8$, $\lambda=1$. }
\label{fig8}
\end{figure}
\begin{figure}[H]
\centering
\subfigure[ $A(r)$, $B(r)$ and $f(r)$ vs $r$]{
\label{9a} 
\includegraphics[width=0.3\textwidth]{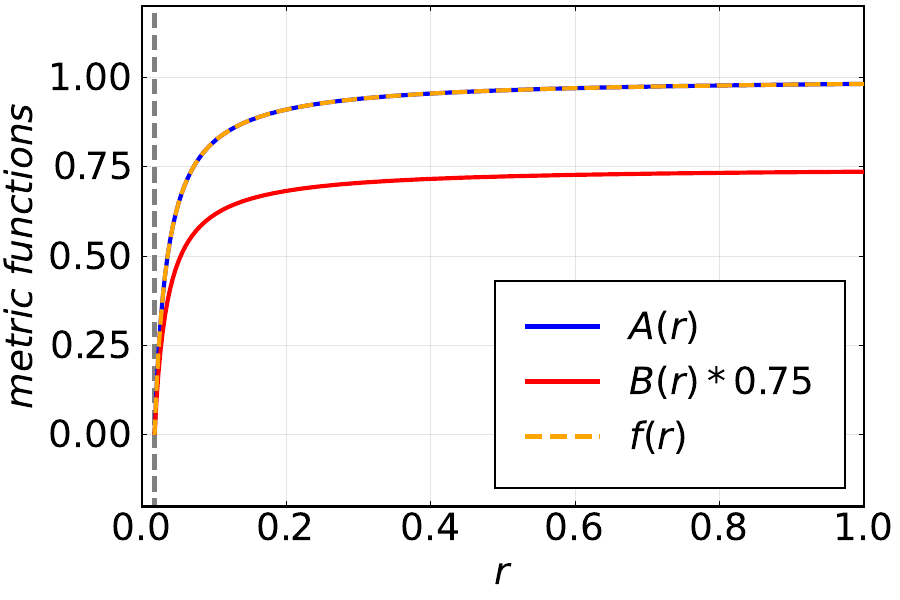}}
\hfill%
\subfigure[ $\delta A(r)$ and $\delta B(r)$ vs $r$]{
\label{9b}
\includegraphics[width=0.3\textwidth]{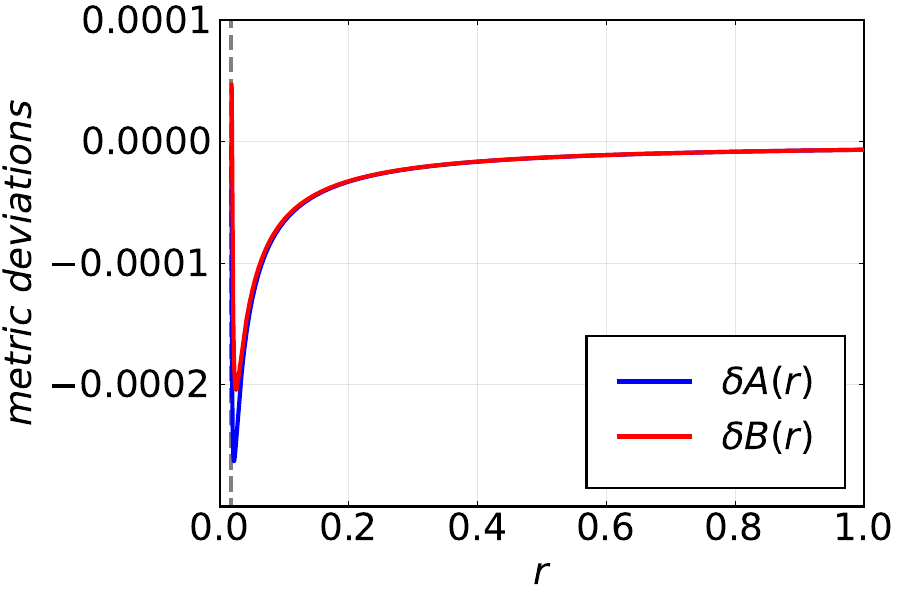}}
\hfill%
\subfigure[$\phi(r)$ vs $r$]{
\label{9c}
\includegraphics[width=0.3\textwidth]{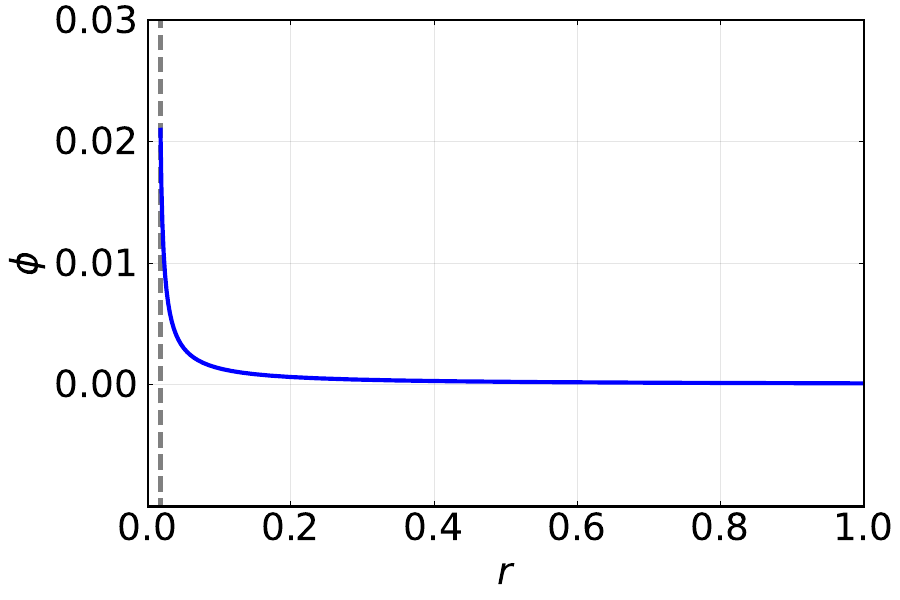}}
\caption{Profile functions of the scalarized black hole on the lower branch and the SBH with $r_H= 0.018$ for the coupling function $\zeta_1(\phi)$ for $\alpha=1/4$, $\beta = 1000/8$, $\lambda=1$.}
\label{fig9}
\end{figure}

Finally, we show two examples for the profile function of the solutions in Figs.~\ref{fig8} and \ref{fig9}.
Here $\zeta_1(\phi)$ is chosen with the coupling parameters $\alpha=1/4$ and $\beta=1000/8$.
Figure \ref{fig8} represents a solution on the primary branch, while Fig.~\ref{fig9} illustrates a solution on the lower branch.
The metric functions $A(r)$ and $B(r)$ are very similar to the metric function $f(r)$ of the SBH, also shown in the figures.
The horizon radius is selected as $r_H= 0.018$. 
The figure shows that the metric functions $A(r)$ and $B(r)$ 
are rather similar to the metric function $f(r)$ of the SBH solution. 
Therefore, we display in Figs.~\ref{8b} and \ref{9b}, the metric deviations $\delta A(r)=A(r)-f(r)$ and $\delta B(r)=B(r)-f(r)$. 
On the lower branch, the deviations of the metric functions from the SBH are in fact extremely small here.
Thus the metric properties are almost those of the SBH, unless a solution close to the maximal $M$ is considered.
The scalar function $\phi(r)$ decreases monotonically with $r$ and approaches zero at infinity as seen in Figs.\ref{8c} and \ref{9c}.
Again, on the lower branch the scalar field is rather small away from the maximal $M$.

\subsection{Comparison with polynomial coupling function $\zeta_2(\phi)$}

\begin{figure}[H]
\centering
\subfigure[$\phi_H$ vs $ M$]{\label{f1f2_2a} 
\includegraphics[width=0.3\textwidth]{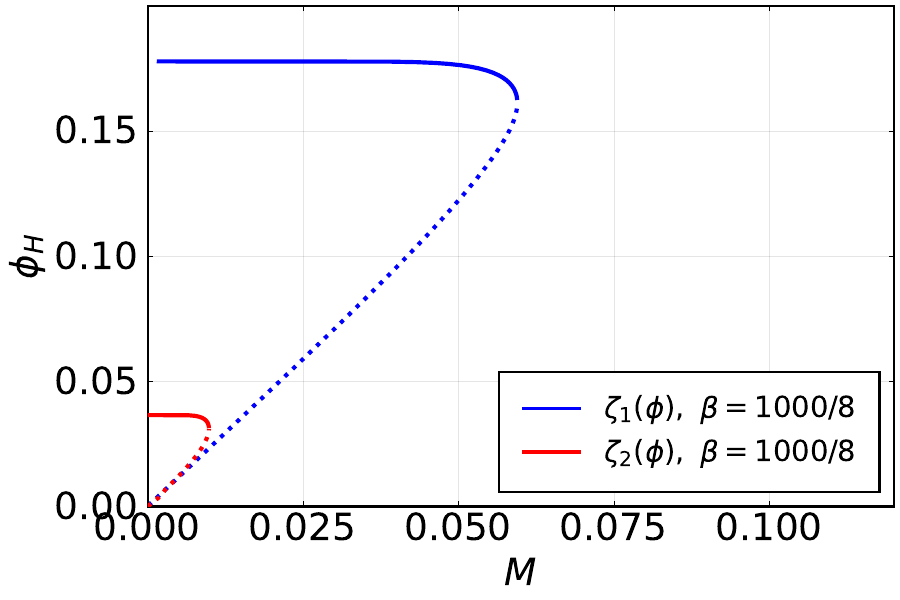}}
\subfigure[$Q_s$ vs $ M$]{
\label{f1f2_2b} 
\includegraphics[width=0.3\textwidth]{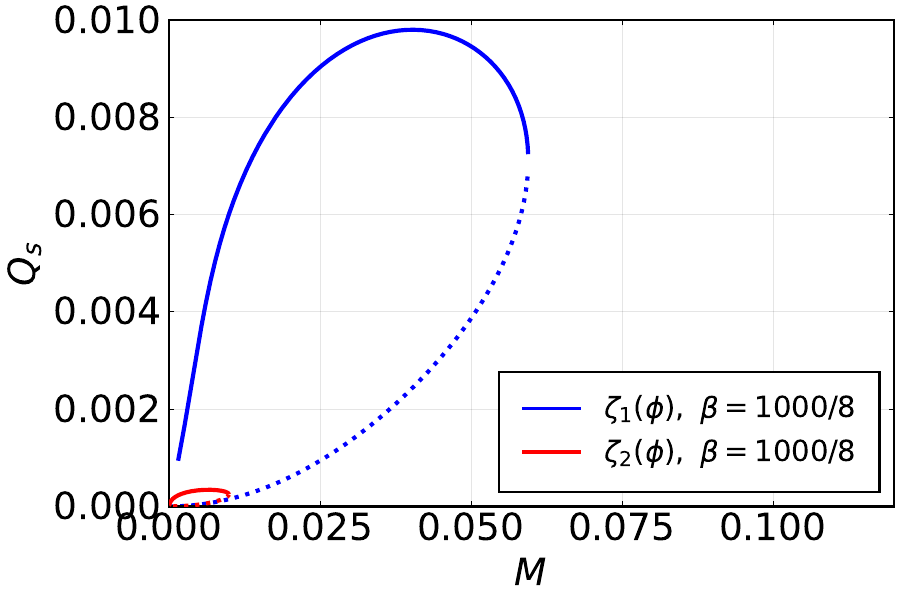}}
\caption{Scalar field $\phi_H$ at the horizon and scalar charge $Q_s$ as functions of the mass $M$ for two coupling functions $\zeta_1(\phi)$ and $\zeta_2(\phi)$ for $\alpha=1/4$, $\beta=1000/8$, $\lambda=1$.}
\label{fig2}
\end{figure}

\begin{figure}[H]
\centering
\subfigure[ $A(r)$, $B(r)$ and $f(r)$ vs $r$]{
\label{9ax}
\includegraphics[width=0.3\textwidth]{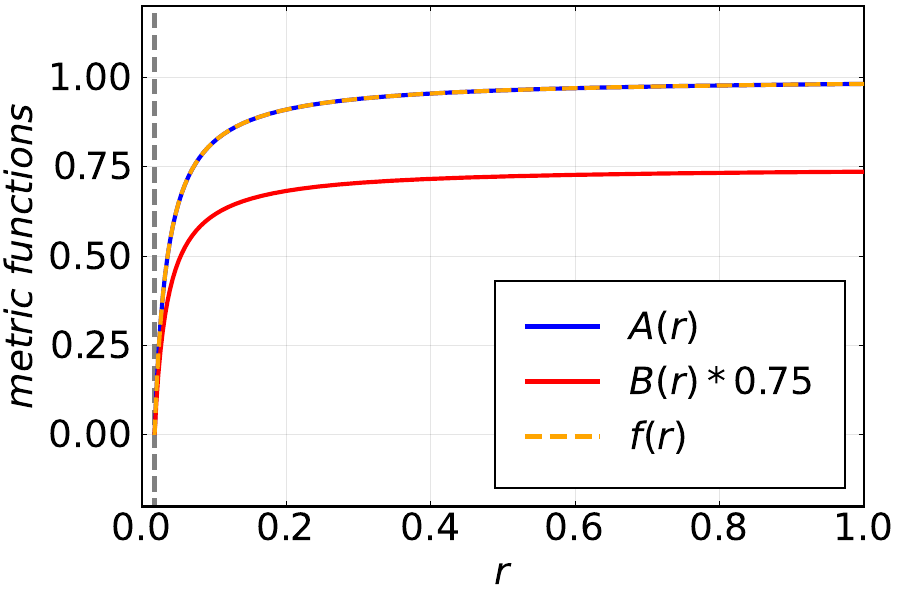}}%
\hfill%
\subfigure[ $\delta A(r)$ and $\delta B(r)$ vs $r$]{
\label{9bx}
\includegraphics[width=0.3\textwidth]{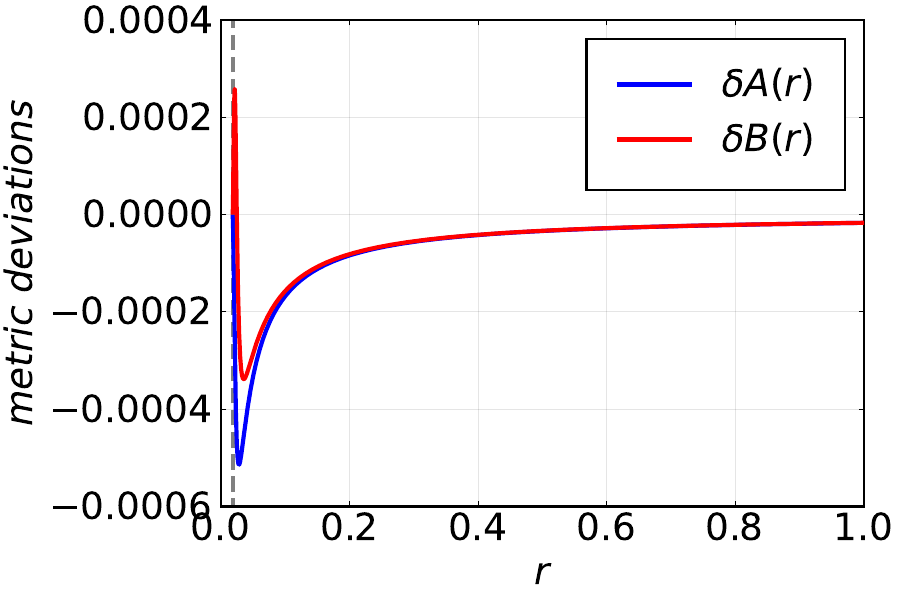}}%
\hfill%
\subfigure[$\phi(r)$ vs $r$]{
\label{9cx}
\includegraphics[width=0.3\textwidth]{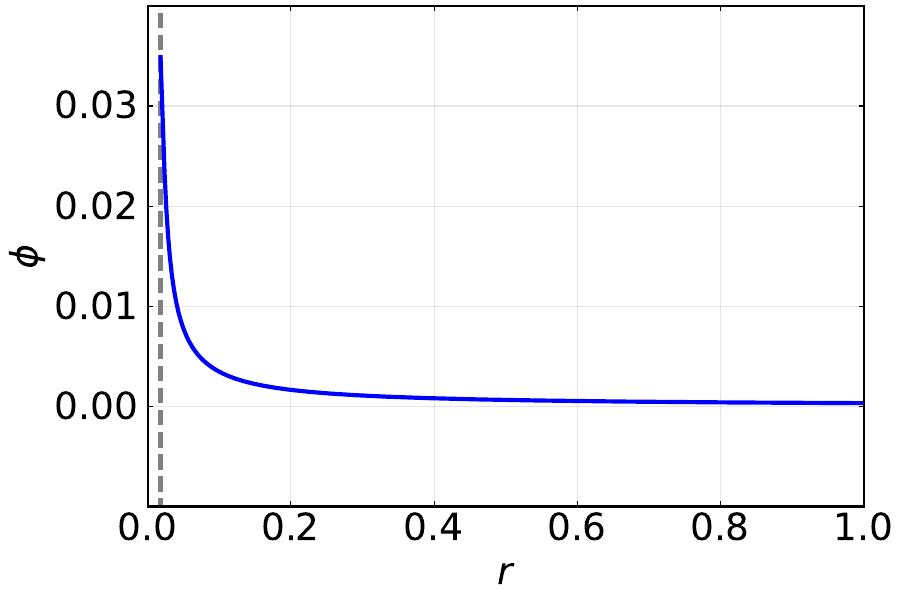}}
\caption{Profile functions of the scalarized black hole and SBH with $r_H= 0.018$ for the coupling function $\zeta_2(\phi)$ for $\alpha=1/4$, $\beta = 1000/8$, $\lambda=1$.}
\label{fig9x}
\end{figure}

The pattern for the solutions for coupling function $\zeta_2(\phi)$ is very similar to the one for $\zeta_1(\phi)$.
Here condition (\ref{const}) implies the existence of a constant $\phi$ solutions for
\begin{eqnarray}
\phi_{\rm s}=\pm\left(\frac{2\alpha}{3\beta}\right)^{1/2}.
\label{phis2}
\end{eqnarray}
We illustrate the solutions for $\alpha=1/4$ and $\beta=1000/8$ ($\lambda=1$) in Fig.~\ref{fig2}, where the solutions for $\zeta_1(\phi)$ are also shown.
For this value of $\beta$, there are two branches.
In the $\phi_H$-$M$ diagram Fig.~\ref{f1f2_2a} the primary branch is associated with $\phi_H=\phi_{\rm s}=0.0365$,
while the lower branch closely follows the linear probe limit curve (for $\beta=0$) until it gets closer to the vicinity of $\phi_{\rm s}$.
The shapes of the $Q_s$-$M$ curves are also analogous for both coupling functions.
Also, the profile functions shown in Fig.~\ref{fig9x} exhibit analogous behavior.


\subsection{Comparison with exponential coupling function}

\begin{figure}[h!]
\centering
\subfigure[$\phi_H$ vs $ M$]{
\label{2a} 
\includegraphics[width=0.3\textwidth]{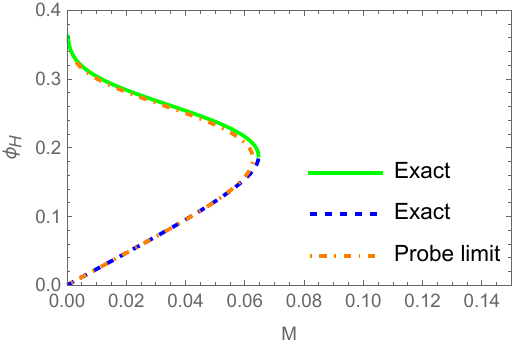}}
\hspace{0.5cm}
\subfigure[$Q_s$ vs $ M$]{
\label{2b} 
\includegraphics[width=0.32\textwidth]{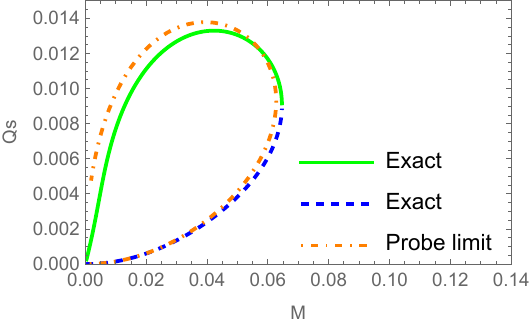}}
\caption{Scalar field $\phi_H$ at the horizon and scalar charge $Q_s$ as functions of the mass $M$ for the exponential coupling function $\zeta_{\rm ex}(\phi) =\frac{1}{4\kappa}\left(1-e^{-\kappa\phi^4}\right)$ with $\kappa=1000/8$.}\label{fig112}
\end{figure}

Nonlinear scalarization with an exponential coupling function was studied \cite{Doneva:2021tvn}-\cite{Blazquez-Salcedo:2022omw}.
Before comparing our results for $\zeta_1(\phi)$ to those for the corresponding exponential coupling function $\zeta_{\rm ex}(\phi)=\frac{1}{4\kappa}\left(1-e^{-\kappa\phi^4}\right)$, we briefly address the probe limit for $\zeta_{\rm ex}(\phi)$.
The probe limit and the exact results for $\zeta_{\rm ex}(\phi)$ are exhibited in Fig.~\ref{fig112} for $\kappa=1000/8$.
Clearly, the probe limit captures the behavior of the solutions for this value of $\kappa$ well.
However, $\frac{d \zeta_{\rm ex}(\phi)}{d \phi}=0$ does not feature a finite solution $\phi_{\rm s}$.
It vanishes only at zero and at infinity.
Thus there are two branches, where the lower branch starts at the origin and then closely follows the linear probe limit curve for $\zeta_3(\phi)$, until the higher order terms in the coupling function take over for sufficiently large $\phi_H$.
But the primary branch is not limited by a finite value of $\phi_{\rm s}$.
Consequently, $\phi_H$ keeps increasing on the primary branch for $M \to 0$.
Because of the exponential coupling function, the scalar field is, however, exponentially suppressed for large $\phi$ in the source terms of the field equations.

\begin{figure}[H]
\centering
\subfigure[$\phi_H$ vs $ M$]{\label{2ay} 
\includegraphics[width=0.3\textwidth]{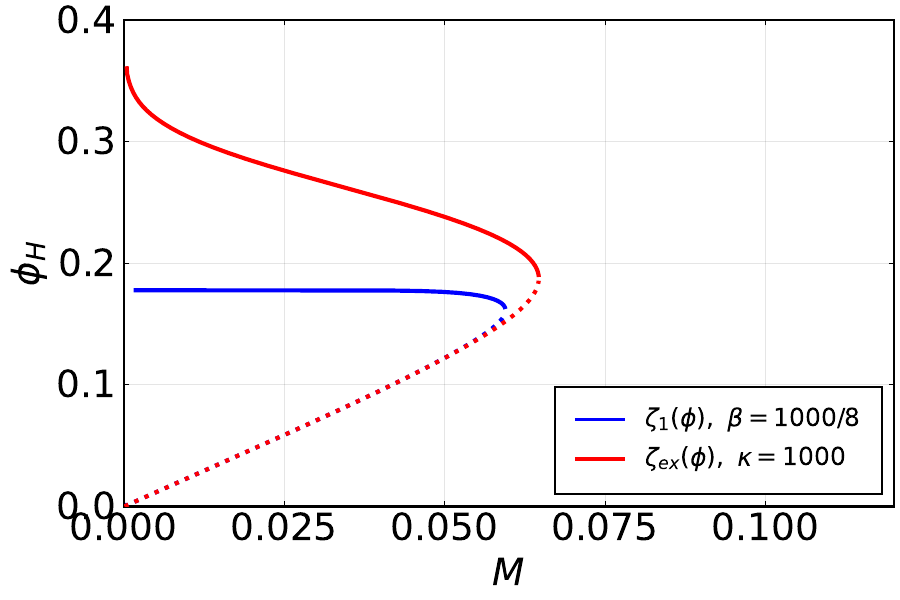}}
\hspace{0.5cm}
\subfigure[$Q_s$ vs $ M$]{
\label{2by} 
\includegraphics[width=0.3\textwidth]{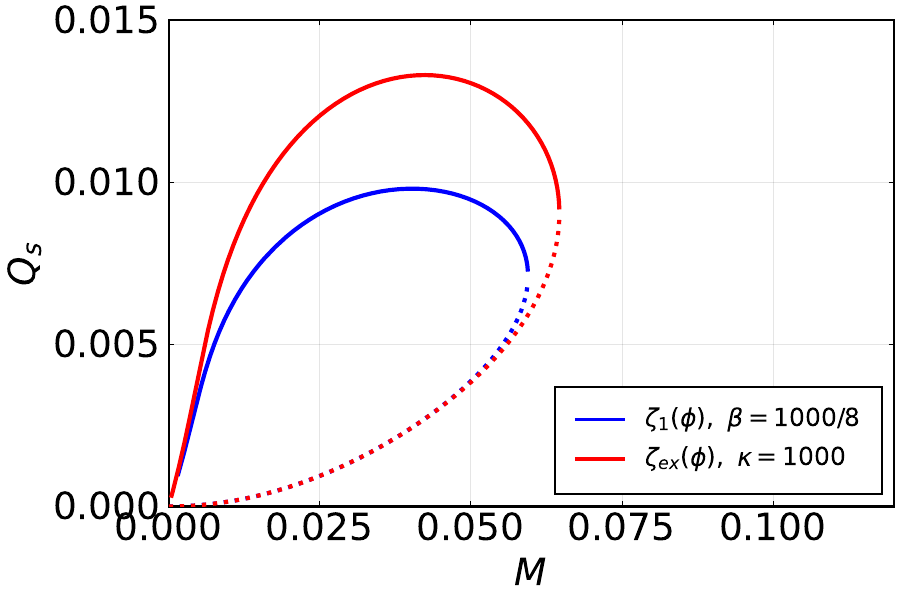}}
\caption{Scalar field $\phi_H$ at the horizon and scalar charge $Q_s$ as functions of the mass $M$ for two coupling functions $\zeta_1(\phi)$ and $\zeta_{\rm ex}(\phi)$ for $\alpha=1/4$, $\beta=\kappa=1000/8$, $\lambda=1$.}
\label{fig2y}
\end{figure}

Figure \ref{fig2y} compares the $\phi_H$-$M$ and $Q_s$-$M$ relations for the exponential coupling function and the corresponding polynomial one for $\beta=\kappa=1000/8$.
As expected, their lower branches largely overlap until the higher nonlinearities become relevant.
The shape of the $Q_s$-$M$ relations remains analogous for both coupling functions, and the pattern of the splitting of the two branches into three branches for smaller values of $\beta=\kappa$ is also equivalent \cite{Blazquez-Salcedo:2022omw}.

\section{Conclusions and discussions}
\label{Sec5}

We investigated nonlinear instability for polynomial coupling functions, that satisfy $\frac{{{d^2}\zeta_i}}{{d{\phi ^2}}}(0) = 0$.
In particular, we considered
$\zeta_1(\phi)= \alpha\phi^4-\beta\phi^8$, $\zeta_2(\phi)=\alpha\phi^4-\beta\phi^6$, and $\zeta_3(\phi)=\alpha\phi^4$. 
Unlike for $\zeta_3(\phi)$, for $\zeta_1(\phi)$ and $\zeta_2(\phi)$, the derivative of the coupling function with respect to the scalar field $\frac{d\zeta_i(\phi)}{d\phi}$ vanishes not only at $\phi=0$, but also at a finite value $\phi_{\rm s}$.
This has interesting consequences for the set of solutions as revealed in the numerical calculations.

First we solved the ($1+1$)-dimensional scalar field equation \eqref{KG2} in the Schwarzschild black hole (SBH) background.
For $\zeta_1(\phi)$ and $\zeta_2(\phi)$, SBHs then 
may transition to scalarized black holes. 
Once the amplitude of the scalar field exceeds its corresponding threshold $A_{\rm th}$,
the scalar field grows first due to the energy carried by the initial perturbation and then, stabilizes to a constant that indicates a formation of scalarized phase. 
In contrast, for $\zeta_3(\phi)$, the scalar field diverges within a short timescale for large initial amplitudes or decays for small initial amplitudes.

Treating the coupling term $\zeta_i(\phi)R^2_{GB}$ as an effective potential $V^i_\text{eff}$, we gained further insight by analyzing the shape of the effective potential. 
For $\zeta_1(\phi)$ (and analogously for $\zeta_2(\phi)$), the $W$-type potential well acts as a trapping region for the scalar field.  
The $\beta\phi^8$ term provides the necessary nonlinear feedback to balance the driving force of the $\alpha\phi^4$ term, forcing the scalar field to settle into a stable stationary state. 
Thus this potential structure provides the explanation for the presence of the ``plateau" observed in the time-evolution results. 
The correspondence between the geometric minimum of $V_\text{eff}$ located at $\phi_{\rm s}$ and the final amplitude of the scalar field confirms that the $\beta$ term is essential for the nonlinear saturation and the existence of (potentially) stable scalarized black hole solutions.

The solutions obtained through time evolution were then recovered by solving the time-independent scalar field equation in the SBH background, i.e., by considering the probe limit or decoupling limit.
The probe limit not only yields solutions for coupling functions $\zeta_1(\phi)$ and $\zeta_2(\phi)$ but also for $\zeta_3(\phi)$.
For the latter coupling function a single branch of solutions is found that is also present for $\zeta_1(\phi)$ and $\zeta_2(\phi)$.
Along this branch, the scalar field $\phi_H$ at the horizon is proportional to the mass $M$.
The coupling functions $\zeta_1(\phi)$ and $\zeta_2(\phi)$ feature, in addition, a second branch, the primary branch, found in the time evolution.
The scalar field $\phi_H$ on the horizon of these solutions remains very close to $\phi_{\rm s}$ for most of the branch.
Both branches merge at a maximal value of the mass, yielding thus a continuous set of solutions.

When taking backreaction into account and solving the coupled set of field equations, the value of the coefficient $\beta$ of the higher order term strongly influences the pattern of the solutions.
These solutions must respect a regularity condition for the scalar field at the horizon.
This condition imposes already for $\zeta_3(\phi)$ an endpoint for its single branch.
Since this branch is rather universal, the double branch pattern of the probe limit can only be retained for large values of $\beta$.
Below a critical value of $\beta$, the set of scalarized solutions features three branches, the lower universal branch, the primary branch and a third intermediate branch that is connected to the primary branch.

When comparing with the corresponding exponential coupling functions $\zeta_{\rm ex}(\phi)$, that possess the polynomial ones as truncations, we find a rather similar picture with respect to the dependence of the branch structure on the coupling constant \cite{Doneva:2021tvn,Blazquez-Salcedo:2022omw}.
Also, the universal lower branches are recovered for $\zeta_{\rm ex}(\phi)$. 
The primary branch differs, however, since it is not associated with a finite $\phi_{\rm s}$.
The scalar field $\phi_H$ at the horizon is not limited by such a constant, but its growth is quenched due to its exponential suppression in the field equations for large values of $\phi$.

While we did not study linear mode stability and hyperbolicity of these scalarized black holes with polynomial coupling functions, we expect that they will have similar such properties as the solutions obtained with exponential coupling functions \cite{Blazquez-Salcedo:2022omw}.
We did study though the thermodynamic stability of the nonlinearly scalarized black holes with polynomial and exponential coupling functions.
Here we found indeed the expected agreement \cite{Herdeiro:2026}.

\vspace{1cm}

{\bf Acknowledgments}

\vspace{1cm}

We would like to thank Zhen-Hao Yang, Chao-Ming Zhang and Lina Zhang for helpful discussions. 
We gratefully acknowledge support by the National Natural Science Foundation of China (NNSFC) (Grant Nos.12365009, 12305064 and 12205123). Y.S.M. was supported by the National Research Foundation of Korea (NRF) grant funded by the Korea government(MSIT) (RS-2022-NR069013).

\end{document}